\documentclass[12pt,preprint]{aastex}

\shorttitle{Dynamics of NGC 5128}
\shortauthors{Woodley et al.}

\begin{document}

\title{The Kinematics and Dynamics of the Globular Clusters and Planetary Nebulae of NGC 5128}

\author{Kristin A.~Woodley}
\affil{Department of Physics \& Astronomy, McMaster University,
  Hamilton ON L8S 4M1, Canada}
\email{woodleka@physics.mcmaster.ca}

\author{William E.~Harris}
\affil{Department of Physics \& Astronomy, McMaster University,
  Hamilton ON L8S 4M1, Canada}
\email{harris@physics.mcmaster.ca}

\author{Michael A.~Beasley}
\affil{Instituto de Astrofisica de Canarias, Calle V\'{i}a L\'{a}ctea, s/n 
E-38200, La Laguna Tenerife, Spain}
\email{beasley@iac.es}

\author{Eric W.~Peng}
\affil{Herzberg Institute of Astrophysics, 5071 West Saanich Road,
  Victoria, BC V8V 2X6, Canada}
\email{eric.peng@nrc-cnrc.gc.ca}

\author{Terry J. Bridges}
\affil{Department of Physics, Queen's University, Kingston, ON K7L 3N6, Canada}
\email{tjb@astro.queensu.ca}

\author{Duncan A.~Forbes}
\affil{Centre for Astrophysics \& Supercomputing, Swinburne University, Hawthorn, VIC 3122, Australia}
\email{dforbes@astro.swin.edu.au}

\author{Gretchen L. H.~Harris}
\affil{Department of Physics \& Astronomy, University of Waterloo, Waterloo ON N2L
  3G1, Canada}
\email{glharris@astro.uwaterloo.ca}

\begin{abstract}
A new kinematic and dynamic study of the halo of the giant elliptical
galaxy NGC 5128 is presented.  From a spectroscopically confirmed
sample of $340$ globular clusters and $780$ planetary nebulae, the
rotation amplitude, rotation axis, velocity dispersion, and 
total dynamical mass are determined for the halo of NGC 5128.  The
globular cluster kinematics were
searched for both radial dependence and metallicity dependence by
subdividing the globular cluster sample
into 158 metal-rich ([Fe/H]$ > -1.0$) and 178 metal-poor ([Fe/H]$ <
-1.0$) globular clusters.  Our results show that the kinematics of the 
metal-rich and metal-poor subpopulations
are quite similar:  over a projected radius of $0-50$ kpc, 
the mean rotation amplitudes are $47\pm15$ and $31\pm14$ km
s$^{-1}$ for the metal-rich and metal-poor populations,
respectively.
There is a indication within $0-5$ kpc that the
metal-poor clusters have a lower rotation signal than in the outer
regions of the
galaxy.  The rotation axis shows an interesting twist at 5 kpc,
agreeing with the zero-velocity curve presented by Peng and coworkers.
Within 5 kpc, both metal-rich and metal-poor populations have a
rotation axis nearly parallel to the north-south direction, which is
$0^o$, while beyond
5 kpc the rotation axis twists $\sim180 ^o$.  The velocity
dispersion displays a steady increase with galactocentric radius
for both metallicity populations, with means of
$111\pm6$ and $117\pm6$
km s$^{-1}$ within a projected radius of 15 kpc for the metal-rich and
metal-poor populations; however, the outermost regions suffer from low
number statistics and spatial biases.  
The planetary nebula kinematics are slightly
different.  Out to a projected radius of $90$
kpc from the center of NGC 5128, the planetary nebulae have a higher rotation
amplitude of $76\pm6$ km s$^{-1}$, and a rotation axis of $170\pm5 ^o$
east of north, with no significant radial deviation in either determined quantity.  The velocity dispersion
decreases with galactocentric distance.  
The total mass of NGC 5128 is found using the tracer mass estimator, described by Evans et
al., to determine the mass supported by internal random
motions and the spherical component of the Jeans equation to determine
the mass supported by rotation.   We find a total mass of 
$1.0\pm0.2 \times 10^{12}$ $M_{\odot}$ from the planetary nebula
data extending to a projected radius of 90 kpc.  The similar kinematics of the metal-rich and 
metal-poor globular clusters allow us to combine the two
subpopulations to determine an independent estimate of the total mass,
giving $1.3\pm0.5 \times 10^{12}$ $M_{\odot}$
out to a projected radius of 50 kpc.  Lastly, we publish a
new and homogeneous catalog of known globular clusters in NGC 5128.
This catalog combines all previous definitive cluster identifications from radial
velocity studies and $HST$ imaging studies, as well as 80 new globular
clusters with radial velocities from a study of M.A. Beasley et
al. (in preparation).
\end{abstract}


\section{Introduction}
\label{sec:intro}

Globular clusters (GCs), as single-age, single-metallicity
objects, are excellent tracers of the formation history of their host
galaxies, through their dynamics, kinematics, metallicities, and ages.
For most galaxies within $\simeq 20$ Mpc, GCs can be
identified through photometry and image morphology, from which 
follow-up radial velocity studies can be carried out with multi-object 
spectroscopy on 4 and 8 m class telescopes.  
The ability to target hundreds of objects in a single field has vastly
increased the observed samples of GCs confirmed in many
galaxies, providing the necessary basis for detailed kinematic and age studies.     

Another benefit of using GCs as a kinematic tracer is that they
provide a useful, independent basis for comparison with results
from planetary nebulae (PNe). This is particularly true given the current debate
surrounding the use of PN velocities and the implications for low
dark matter halos.  For example, \cite{romanowsky03} reported, based on PN
velocities, that three low-luminosity ellipticals
revealed declining velocity dispersion profiles and little
or no dark matter.  However, subsequent simulations of merger-remnant ellipticals suggested that the radial anisotropy of
intermediate-age PNe could give rise to the observed profiles
within standard halos of dark matter \citep[][]{dekel05,mamon05}. 
Flattening of the galaxy along the line of sight is
another possible explanation.  One of the ellipticals studied by
\cite{romanowsky03}, NGC 3379, has also been investigated using GC kinematics
\citep[][]{pierce06,bergond06}; both studies
found evidence of a dark matter halo.  But the two studies of NGC 3379
suffered from small number statistics.  Clearly there is a need to
directly compare PN and GC kinematics for the same elliptical
with sufficiently large numbers of tracer objects.

Previous velocity-based studies of globular cluster systems (GCSs) have shown
an intriguing variety of
results in their overall kinematics.  
Prominent recent examples include the following: 

1.  \cite{cote01} performed a kinematic analysis of the GCS in M87 (NGC 4486), the cD
galaxy in the Virgo Cluster.  With a sample of $280$ GCs,
they showed that the entire GCS rotates on an 
axis matching the photometric
{\it minor} axis of the galaxy, except for the inner metal-poor
sample.  Inside the onset of the cD envelope, the
metal-poor clusters appear to rotate around the {\it major} axis of the
galaxy instead.  This study also found evidence for an increase in
velocity dispersion, $\sigma_v$, with radius, due to the larger scale
Virgo Cluster mass distribution.   
They also showed no strong evidence for a difference in $\sigma_v$ 
between the metal-poor and metal-rich groups.  Using a Virgo mass
model, they investigated anisotropy and found that as a whole, the GCS
had isotropy, but considered separately, the metal-poor and
metal-rich subpopulations had slight anisotropy.

2.  \cite{cote03} performed a dynamical analysis for M49 (NGC 4472),
the other supergiant member of the Virgo Cluster.  Using over $260$
GCs, they found that the metal-rich population shows no
strong evidence for rotation, while the metal-poor population does
rotate about the minor axis of the galaxy.  In addition, they 
found that the metal-poor clusters had an overall higher
dispersion than the metal-rich population.

3.   \cite{richtler04} determined the kinematics for the GCS 
in NGC 1399, the brightest
elliptical in the Fornax Cluster of galaxies.  Armed with a sample 
of over $460$ GCs, they found a marginal
rotation signal for the entire GC sample and the outer
metal-poor sample, while no rotation was seen for the metal-rich
subpopulation.  Their projected velocity dispersion showed no radial
trend within their determined uncertainty, but the metal-poor
clusters had a higher dispersion than the metal-rich clusters.  The
kinematics of the PNe in NGC 1399 have been most recently studied by
\cite{saglia00} and \cite{napolitano02} using a small sample of 37 PNe
from \cite{arnaboldi94}.  These studies do not indicate any
significant deviations in velocity dispersion with radius for the PNe, yet
interestingly, \cite{napolitano02} found stong rotation for the PNe 
in the inner region of the galaxy.  

4.   \cite{pff04II} studied the GC kinematics of the giant elliptical NGC
5128 from a total of $215$ GCs.  Their results showed
definite rotation signals in both metallicity groups beyond a 5 kpc
distance from the center of the galaxy, as well as similar velocity
dispersions in both the metal-poor and the metal-rich populations
within a 20 kpc projected radius.  A similar PN kinematic study in NGC 5128 
\citep{peng04} showed that the PNe population is rotating around a 
twisted axis that turns just beyond the 5 kpc distance.

These few detailed kinematic studies of GCSs in
elliptical galaxies, show results that appear to
differ on a galaxy-by-galaxy basis, without any clear global trends.
The age distributions of
the GC populations in these same galaxies tend to show a
consistent pattern in which the blue, or metal-poor, population
is found to be universally old.  The red, or metal-rich,
population has also been shown to be old in the study of
\cite{strader05}.  Their study found that {\it both} the metal-poor and metal-rich GCs in a sample of eight
galaxies ranging from dwarf to massive ellipticals, all have ages as old
as their Galactic GC counterparts.  Conversely, in a small sample of
studies, the red GC population has also been found to be 2-4 Gyr
younger than the metal-poor GC population, and with a wider spread in determined
ages \citep[see the studies of][among others]{pff04II,puzia05},
although this is not yet a well-established trend.  The GCs in NGC
5128 appear to be old, with an intermediate-age
population \citep{pff04II,beasley06}. 

NGC 5128 (Centaurus A), the central giant in the Centaurus group of
galaxies at a distance of only $\sim4$ Mpc, is a prime candidate for both kinematic
and age studies.  Its GCS has a specific frequency of $S_N \simeq 2.2\pm0.6$
\citep{harris06}, toward the low end of the giant elliptical range
but about twice as high as in typical disk galaxies.
Its optical features show faint
isophotal shells located in the halo \citep[e.g.][]{peng02}, the
prominent dust lane in the inner 5 kpc, the presence of gas, and star
formation, all of which suggest that NGC 5128 could be a merger product.
\cite{baade54} first suggested that NGC 5128 could be the result of a
merger between two galaxies, a spiral and an elliptical.  This idea
was followed by the
general formation mechanism of disk-disk mergers proposed by \cite{toomre72}.
\cite{bhh03} found that the metallicity
distribution function of the halo field stars could be reproduced by
a gas-free (''dry'')
disk-disk merger scenario.  Recent numerical simulations by
\cite{bekki06} also demonstrate that the PN kinematics observed in NGC 5128 \citep{peng04} can be
reproduced relatively well from a merger of unequal-mass disk galaxies (with one galaxy
half the mass of the other) colliding on a highly inclined orbital
configuration.   

Alternatively, much of NGC 5128 could be a ''red and dead''
galaxy, passively evolving since its initial formation as a large seed
galaxy \citep{w06}, while undergoing later minor mergers and
satellite accretions.  Evidence consistent with this scenario is in the halo population 
of stars in NGC 5128, which have a mean age of $8^{+3.0}_{-3.5}$ Gyr \citep{rejkuba05}.  
Its metal-poor GC ages have also been shown to have ages
similar to Milky Way GCs, while the
metal-rich population appears younger \citep{pff04II}.  Our new
spectroscopic study 
\citep{beasley06} suggests that NGC 5128 
has a trimodal distribution of cluster ages:  $\sim50\%$ of metal-rich clusters 
have ages of $6-8$ Gyr, only a small handful of metal-rich
clusters have ages of $1-3$ Gyr, and a large fraction of both
metal-rich and metal-poor clusters have ages of $\simeq 12$ Gyr.
Lastly, the halo kinematics of NGC 5128 have also been recently shown to match the
surrounding satellite galaxies in the low-density Centaurus group,
suggesting that NGC 5128 acts like an inner component to its galaxy
group \citep{w06}.  Kinematic and age studies with large number of GCs 
are thus starting to help disentangle the formation of this giant
elliptical. 

The confirmed GC population in NGC 5128 is now large
enough to allow a new kinematic analysis
subdivided to explore radial and metallicity dependence, while avoiding small number statistics
in almost all regions of the galaxy.
The analysis presented here complements 
the detailed age distribution study provided by \cite{beasley06}.  The
results provide a broader picture of the formation scenario of NGC 5128.  

The sections of this paper are divided as follows:
\S~\ref{sec:cat} contains the full catalog of NGC 5128 GCs with known photometry and radial velocities,
\S~\ref{sec:kin} contains the kinematic analysis of GCSs, \S~\ref{sec:kin_PN} contains the kinematic analysis of
the PN population, \S~\ref{sec:dyn} 
contains the discussion of the dynamical mass of NGC 5128, and
\S~\ref{sec:concl} contains our final discussion, as well as
concluding remarks.

\section{The Catalog of Globular Clusters in NGC 5128}
\label{sec:cat}

Finding GCs in NGC 5128 is challenging.  This process begins with photometric surveys
of the many thousands of objects projected onto the NGC 5128
field \citep[e.g.][]{rejkuba01,hhg04II}; only a few percent of these
are the GCs that we seek.  This daunting task is made
difficult by the galactic latitude of NGC 5128 ($b=19^o$), which means
that many foreground stars are present in the field of
NGC 5128.  Background galaxies are another major contaminant in the
field, forcing the use of search criteria such as magnitude, color, and object
morphology to help build the necessary candidate list of GCs. 
Confirmation of such candidate objects can then be done with
spectroscopic radial velocity measurements.
The GCs in NGC 5128 have
radial velocities in the range $v_r = 200-1000$ km s$^{-1}$, while
most foreground
stars have $v_r < 200$ km s$^{-1}$.  Background galaxies have radial velocities of
many thousands of km s$^{-1}$ and can easily be 
eliminated \citep[see the recent studies of][]{pff04I,whh05}. 

Over the past quarter century, there have been seven distinct radial
velocity studies to identify GCs in NGC 5128
\citep{vhh81,hhvh84,hhh86,hghh92,pff04I,whh05,beasley06}.  Although
\cite{hghh92} was not a radial velocity study itself, but rather a CCD
photometric study of previously confirmed GCs, it does include the GCs
determined spectroscopically from \cite{sharples88}.  A recent 
study with measured GC radial velocities published by \cite{rejkuba07}  
has confirmed two new GCs, HCH15 and R122, included in our catalog.

Within this
catalog are 80 new GCs with radial velocity measurements from
\cite{beasley06}.
The combination of these studies now leads to a confirmed population of
342 GCs.\footnote{The confirmed GC list in \cite{w06} containing
  343 GCs has been reduced to 340 based on recent spectroscopic and imaging
  studies.  Object 304867 with a high radial
velocity of $305\pm56$ km s$^{-1}$ appears
to be an M-type star based on the strong molecular bands in its spectrum, see
\cite{beasley06} for further discussion.  Objects pff\_gc-010 and
114993 are also rejected as GCs because of their
starlike appearance under $HST$ ACS imaging; see \cite{harris06}.  
However, newly confirmed GCs HCH15 and R122 have now been added \citep{rejkuba07}.}
Also included in our catalog are the new GC
candidates from $HST$ STIS imaging from \cite{hhhm02}, labelled
C100-C106, and from $HST$ ACS imaging from \cite{harris06}, labelled
C111-C179.   All these previous studies have their own internal
numbering systems, which makes the cluster identifications somewhat
confusing at this point.  Here we define a new, homogeneous listing
combining all this material and with a single numbering system.  

Our catalog of the GCs of NGC 5128 is given in 
Table~\ref{tab:cat_GC}.  In successive columns, the
Table gives the new cluster name in order of increasing right ascension;
the previous names of the cluster in the literature;
right ascension and declination (J2000); the projected radius from the center of
NGC 5128 in arcminutes; 
the $U$, $B$, $V$, $R$, and $I$ photometric indices and their 
measured uncertainties; the $C$, $M$, and $T_1$ photometric indices and their
uncertainties; the colors $U-B$, $B-V$, $V-R$, $V-I$, $M-T_1$, $C-M$,
and $C-T_1$; and, lastly, the weighted mean velocity $v_r$ and its
associated uncertainty from all previous studies.  All $UBVRI$
photometry is from the imaging survey described in  \cite{pff04I}.  
The $CMT_1$ data are from \cite{hhg04II}.

The mean velocities are weighted averages with weights on each
individual measurement equal to
$\varepsilon_{v}^{-2}$ where $\varepsilon_{v}$ is the quoted velocity uncertainty
from each study.  The uncertainty in the mean velocity is then
$<\varepsilon_v> = (\sum \varepsilon_{i}^{-2})^{-1/2}$.
There are no individual uncertainties supplied for the velocities for
clusters studied by \cite{hhh86}, but their study reports that the mean
velocity uncertainty for clusters with $R_{gc} < 11'$ is 25 km
s$^{-1}$ and for $R_{gc} > 11'$ is 44 km s$^{-1}$.  We have adopted
these values accordingly for their clusters.\footnote{\cite{hhh86}
report C32 at a distance of $R_{gc} = 10.8'$, but more
recently \cite{pff04I} claim a distance of $R_{gc} =
11.25'$, so it has an adopted uncertainty of 44 km s$^{-1}$ in the
weighted mean.}  

The study by \cite{hghh92} also does not report velocity
uncertainties; however, these clusters have all
been recently measured by \cite{pff04I}.  The rms scatter of the
\cite{hghh92} values from theirs was 58 km s$^{-1}$.  This value has been adopted as the velocity
uncertainty of the \cite{hghh92} clusters in the weighted means.

The weighted mean velocity of cluster C10 does not include the
measured value determined by \cite{pff04I} which is significantly
different from other measurements.  Also, cluster C27 does not include
the measurement of $v_r = 1932\pm203$ km s$^{-1}$ from \cite{beasley06},
indicating that this object is a galaxy.  We include C27 as a GC, but with caution.  

In the weighted velocity calculations, the velocities and
uncertainties of the 27 GCs from \cite{rejkuba07} have been rounded to the
nearest whole number, with any velocity uncertainty below 1 km
s$^{-1}$ rounded up to a value of 1.

Lastly, the GC pff\_gc-089
overlaps the previously existing confirmed cluster, C49, within a 0.5''
radius;  pff\_gc-089 is therefore removed from the catalog of
confirmed GCs. 

The data in Table~\ref{tab:cat_GC} provide the basis for
the kinematic study presented in this paper.  We use them to derive the rotation amplitude,
rotation axis and velocity dispersion in the full catalog of clusters, as well
as for the metal-poor ([Fe/H] $< -1$) and metal-rich ([Fe/H] $> -1$) 
subpopulations.  For this purpose, we define the metallicity of the
GCs by transforming the dereddened
colors $(C - T_1)_o$ to [Fe/H] through the standard conversion 
\citep{harris02}, calibrated through Milky Way cluster
data.  A foreground reddening value of $E(B - V) =
0.11$ for NGC 5128, corresponding to $E(C - T_1) = 0.22$, has been
adopted.  The division of [Fe/H] = -1 between metal-rich and
metal-poor GCs has been shown as a good split between the two
metallicity populations from [Fe/H] values converted from $C - T_1$ 
in \cite{whh05} and \cite{hhg04II} for NGC 5128.
If no $C$ and/or $T_1$ values are available for the cluster, it is classified as
metal-rich or metal-poor through a transformation from $(U-B)_o$ to
[Fe/H] from \cite{reed94}.

\begin{figure}
\plotone{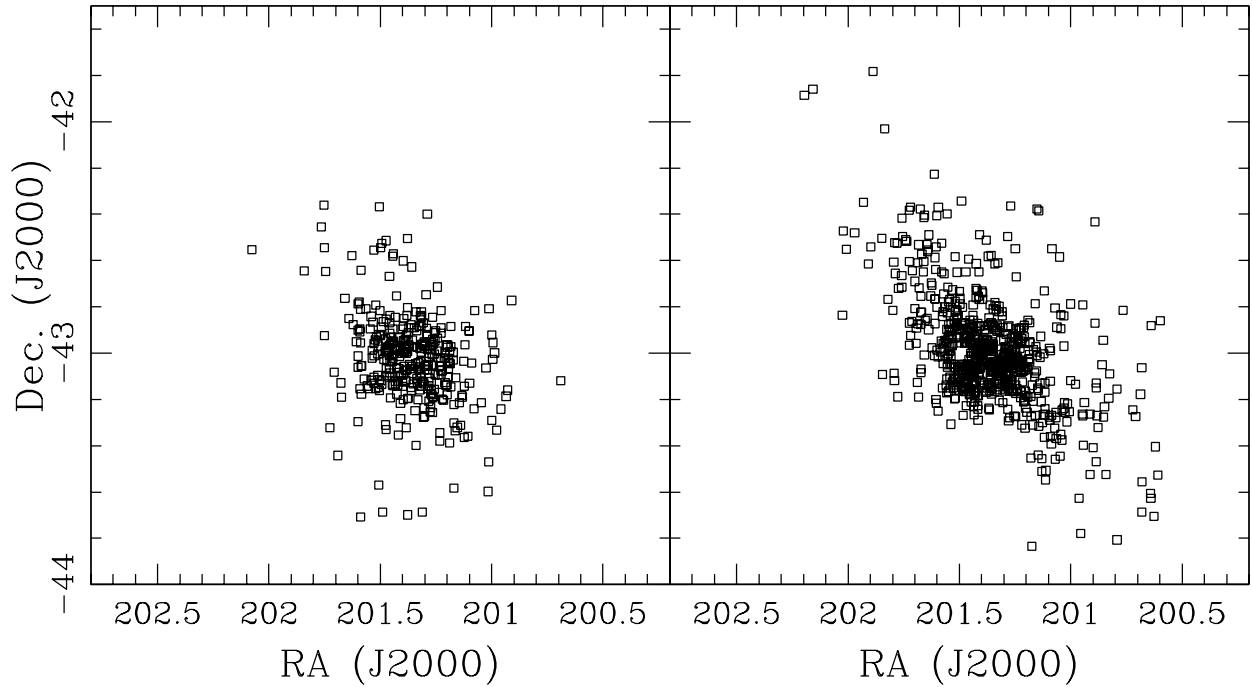}
\caption{Positions of the GCS ({\it left}) totaling 340 objects and the PN system
    ({\it right}) totaling 780 objects in NGC 5128.  The PN
    system has more than double the objects and extends $\sim 40$ kpc
    further out from the center of NGC 5128.  Both systems are
    spatially biased to the major axis of the galaxy with a position
    angle of $35^{o}$ east of north.} 
\label{fig:position}
\end{figure}

In Figure~\ref{fig:position} we show the spatial distributions of all the
GCs from Table~\ref{tab:cat_GC} ({\it left}) and the
distribution of the known PNe ({\it right}).  Both
systems are spatially biased to the major axis of the galaxy because
this is where most of the GC and PN searches have concentrated.

\section{Kinematics of the Globular Cluster System}
\label{sec:kin}

\subsection{Velocity Field}
\label{sec:velfield}

For the present discussion we adopt a distance of $3.9$ Mpc
for NGC 5128. This value is based on four stellar standard
candles that each have internal precisions near $\pm 0.2$ mag:
the PN luminosity function,
the tip of the old-halo red giant branch,
the long-period variables, and the Cepheids
\citep{hhp99,rejkuba04,ferr06}.

HCH15 and R122 have not been included in our kinematic study, as our study was 
completed before publication of these velocities.  The weighted velocities 
used in this kinematic study do not include the most recent 25 velocities 
of previously known GCs published in \cite{rejkuba07}.  Note that the velocities 
published in Table~\ref{tab:cat_GC} do, however, include the \cite{rejkuba07} 
velocities in the quoted final weighted radial velocities for completeness.

\begin{figure}
\plotone{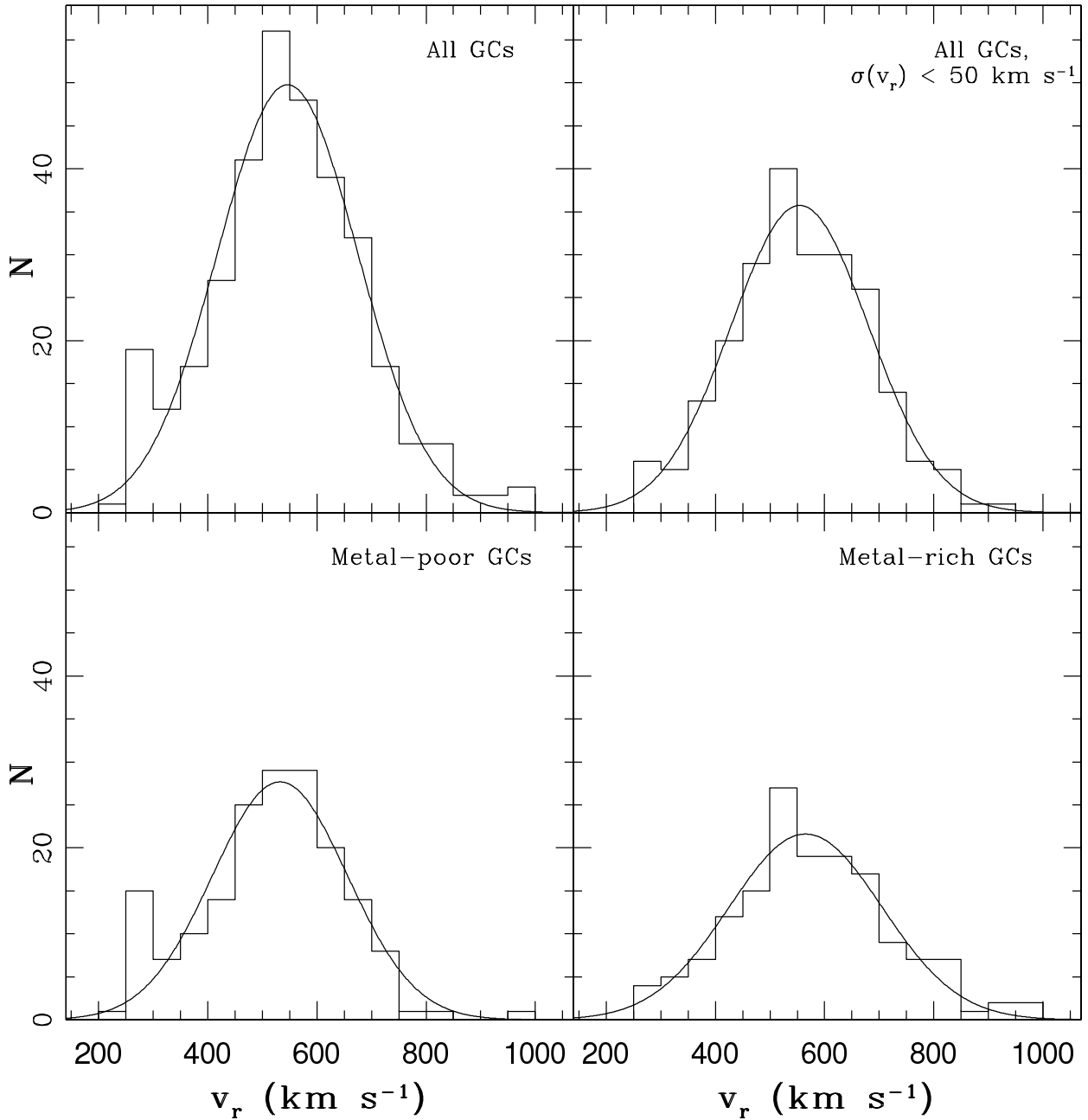}
\caption{Velocity histograms fit with a Gaussian (mean velocity and
  standard deviation in km s$^{-1}$) for the samples. 
  {\it Top left}, the entire GC ($546.4\pm7.3$, $128.8\pm7.3$); {\it top right},
  GCs with a radial velocity
  uncertainty $< 50$ km s$^{-1}$ ($554.8\pm5.3$, $126.2\pm7.3$); {\it
  bottom left}, all metal-poor GCs
  ($532.5\pm12.7$, $123.8\pm12.8$); {\it bottom right}, all metal-rich GCs ($565.5\pm10.5$, $140.3\pm10.7$).} 
\label{fig:gausfit_all}
\end{figure}

The velocity distribution of the entire sample of 340 is
shown in Figure~\ref{fig:gausfit_all} ({\it top left}), binned
in 50 km s$^{-1}$ intervals.  A fit
with a single Gaussian yields a mean velocity of $546\pm7$ km
s$^{-1}$, nicely matching the known systemic velocity of $541\pm7$ km
s$^{-1}$ \citep{hui95}.  There is a slight
asymmetry at the low-velocity end that is likely due to contamination by
a few metal-poor
Milky Way halo stars (also seen in the metal-poor subpopulation in the
bottom left panel, which has a mean velocity determined
by the Gaussian fit as $532\pm13$ km s$^{-1}$).  

Selecting the clusters with radial velocity
uncertainties less than 50 km s$^{-1}$ leaves 226 clusters, plotted in
Fig.~\ref{fig:gausfit_all} ({\it top right}).  The close fit
to a single Gaussian is consistent with an
isotropic distribution of orbits; the mean velocity is $554\pm5$ km
s$^{-1}$.  The metal-rich population, with a mean velocity determined
by the Gaussian fit of $565\pm11$ km s$^{-1}$, is plotted in the 
bottom right panel, and also shows no strong asymmetries.

\begin{figure}
\plotone{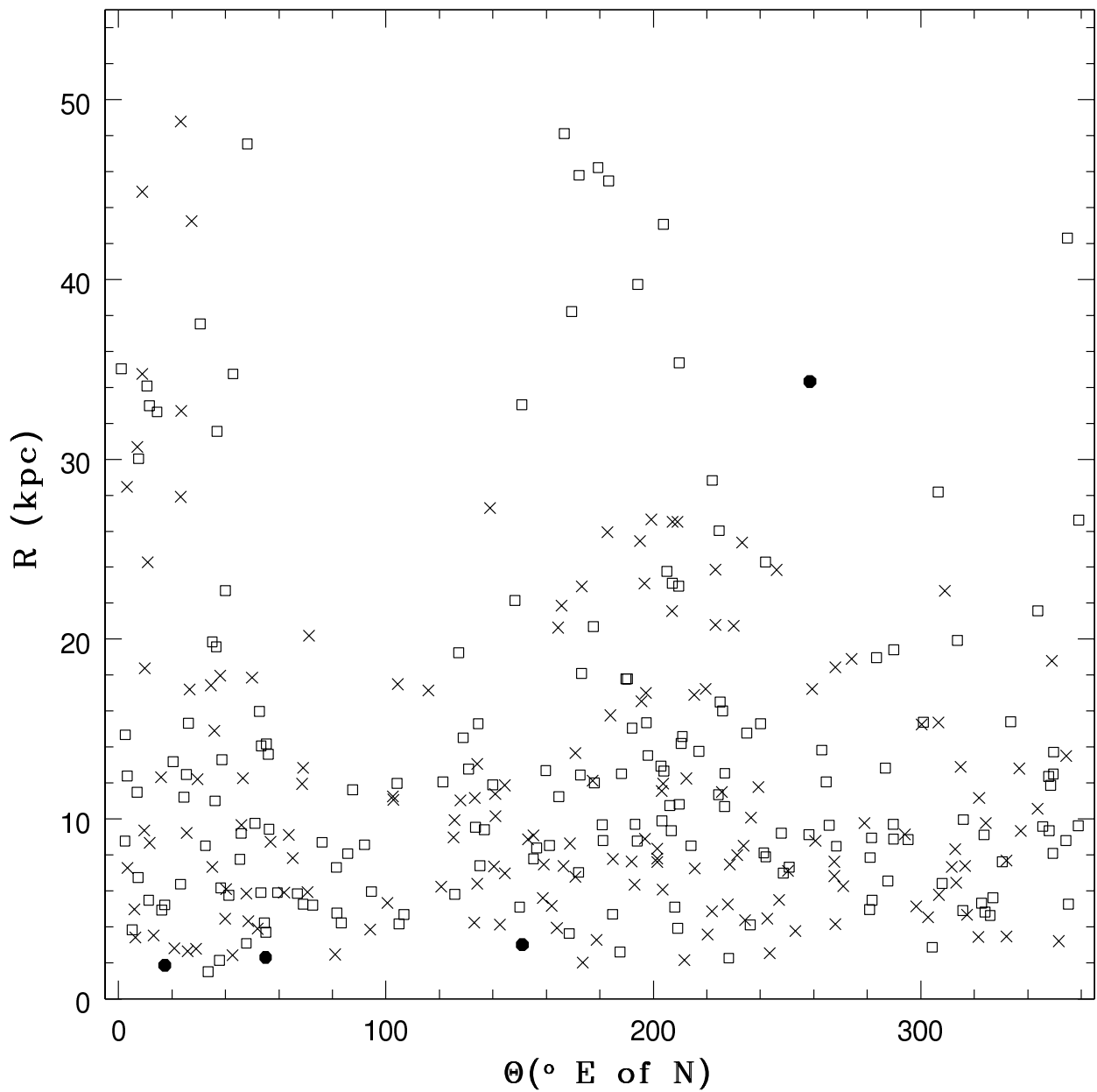}
\caption{Projected radial distribution as a function of projected
  angular position of all known GCs in NGC 5128.  The
  angular distribution of metal-poor ({\it squares}), metal-rich
  ({\it crosses}), and unknown metallicity ({\it circles})
  subpopulations of GCs clearly shows the observational bias in
  GC searches along the photometric minor axis \citep[$\Theta
  = 119$ and $299^{o}$ east of north, ][]{dufour79} beyond 15 kpc.} 
\label{fig:gc_thetar}
\end{figure}

Looking closer at the metal-poor velocity asymmetry, we note that the 15 metal-poor 
clusters between 250 and 300 km s$^{-1}$ (in the region where
contamination by Milky Way field stars could occur) are balanced by
only two GCs at the high-velocity end on reflection across the systemic velocity.   The same
velocity regions in the metal-rich population are nearly equally
balanced with four clusters between 250 and 300 km s$^{-1}$ with three
clusters at the reflected high-velocity range.  Interestingly, the four metal-rich
clusters between 250 and 300 km s$^{-1}$ have projected radii $> 17$ kpc
even though the metal-rich population is more centrally concentrated
than the metal-poor \citep[see][among others]{pff04II,whh05}.  The
metal-poor clusters between 250 and 300 km s$^{-1}$, conversely, are more
evenly distributed, with five clusters
between projected radii of 5 and 10 kpc, five clusters between 10 and 20 kpc,
and five clusters beyond 20 kpc from the center of NGC 5128.  
Some of these low-velocity, metal-poor objects could be foreground 
stars with velocities in the realm
of GCs in NGC 5128 ($v_r \gtrsim 250$ km s$^{-1}$).
However, with only 340 GCs currently confirmed within $\sim45$'
from the center of NGC 5128, out of an estimated $\simeq
1500$ total 
clusters within 25' \citep{harris06}, these metal-poor, low-velocity objects
could simply be part of a very incomplete GC sample that is also spatially biased.  This potential bias is clearly shown in 
Figure~\ref{fig:gc_thetar}, which shows the projected radial distribution as a function of azimuthal
angle for our GC sample.  Beyond 12 kpc, the two ''voids'' 
coincide with the photometric minor axis of
the galaxy, attributed at least partly to incomplete cluster surveys in
these regions.  These objects should, therefore,
not be dropped from the GC catalog without further
spectroscopic analysis.  

\subsection{Rotation Amplitude, Rotation Axis, and Velocity Dispersion}
\label{sec:kin_gc}

\subsubsection{Mathematic and Analytic Description}
\label{sec:math}

We determine the rotation amplitude and axis of the GCS of
NGC 5128 from
\begin{equation}
\label{eqn:kin}
v_r(\Theta) = v_{sys} + \Omega R  sin(\Theta - \Theta_o)
\end{equation}
\citep[see][]{cote01,richtler04,w06}.
In Equation~\ref{eqn:kin}, $v_r$ is the observed radial velocity of the GCs in the
system, $v_{sys}$ is the galaxy's systemic velocity, $R$ is the
projected radial distance of each GC from the center of the
system assuming a distance of 3.9 Mpc to NGC 5128, and $\Theta$ is the projected azimuthal angle of the GC measured
in degrees east of north.  The systemic velocity of NGC 5128 is held
constant at $v_{sys}=541$ km s$^{-1}$ \citep{hui95}
for all kinematic calculations.   The rotation axis of
the GCs, $\Theta_o$, and the product $\Omega R$, the rotation amplitude of the
GCs in the system, are the values obtained from the
numerical solution.  We use a Marquardt-Levenberg non-linear fitting
routine \citep{press92}.  

Eqn.~\ref{eqn:kin} assumes spherical
symmetry.  While this may be a decent assumption for the
inner 12 kpc region \citep[it has a low ellipticity of
  $\sim0.2$;][]{peng04}, true ellipticities for the outer regions of
the system are not well known because of the sample bias (see
Fig.~\ref{fig:gc_thetar}).   
Future studies to remove these biases are vital to obtaining a sound
kinematic solution for the entire system.  Eqn.~\ref{eqn:kin} also
assumes that $\Omega$ is only a function of the projected radius and that the rotation axis
lies in the plane of the sky.  It is not entirely clear how these assumptions, discussed thoroughly
in \cite{cote01}, apply to the GC and PN systems of NGC 5128.
The $\Omega$ we solve for is, therefore, only a lower limit to the
true $\Omega$ if the true rotation axis is not in the plane of the
sky.  

The projected velocity dispersion is also calculated from the normal condition,
\begin{equation}
\label{eqn:veldisp}
\sigma_{v}^{2} =  \sum_{i=1}^{N}\frac{(v_{f_i} - v_{sys})^2}{N}
\end{equation}
where $N$ is the number of clusters in the sample, $v_{f_i}$ is the
GC's radial velocity {\it after subtraction of the rotational
component determined with Eqn.~\ref{eqn:kin}}, and $\sigma_v$ is the projected velocity dispersion.  

\begin{figure}
\begin{center}
\includegraphics[scale=0.95]{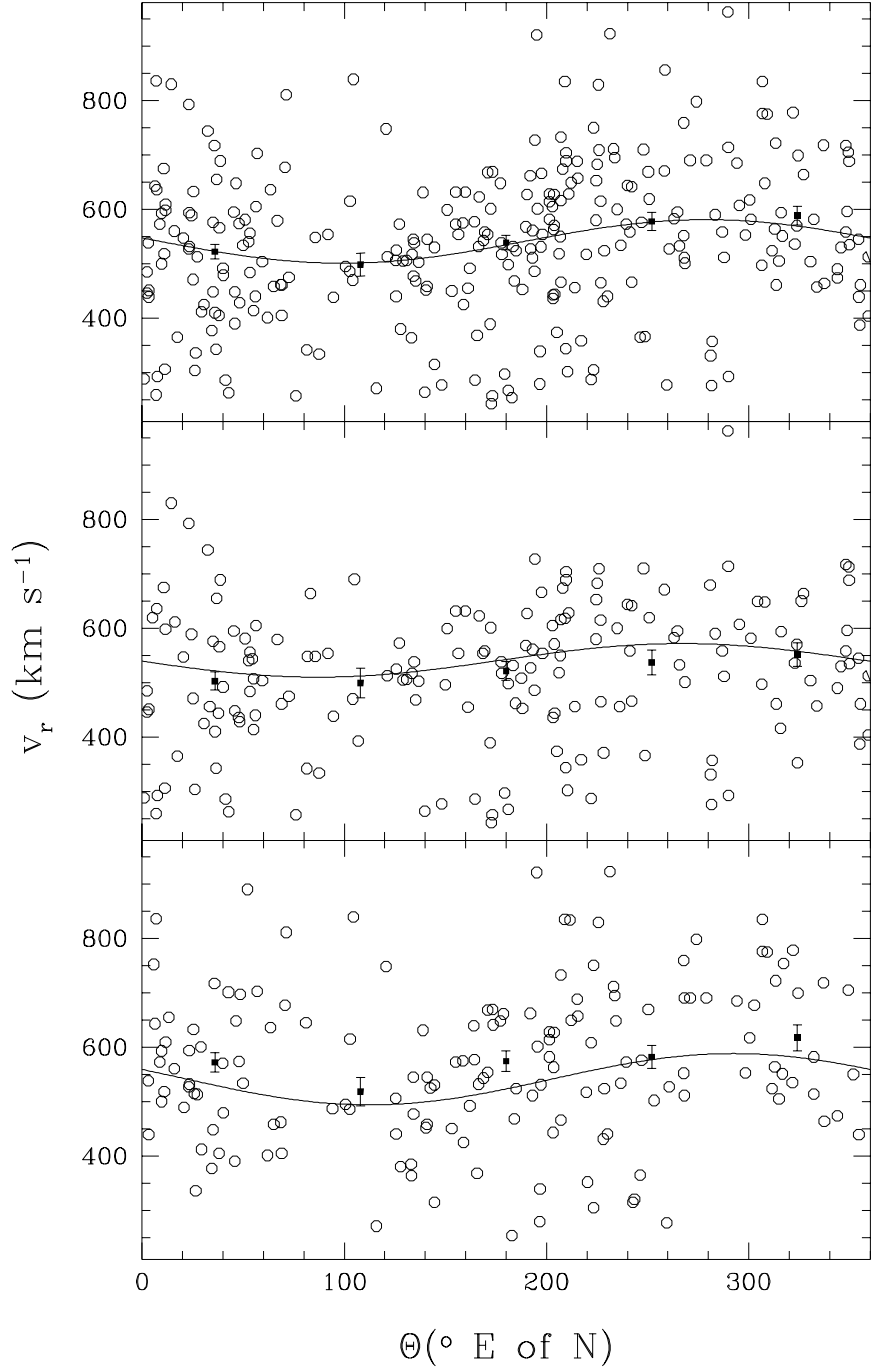}
\caption{Sine curve fit for the GCs in NGC 5128
  ({\it circles}) with a fixed systemic velocity
  of $v_{sys} = 541$ km s$^{-1}$, for 0-50 kpc from the center of NGC
  5128.  The top panel shows all 340 GCs with
  rotation amplitude $\Omega R = 40\pm10 $ km s$^{-1}$ and rotation axis
  $\Theta_o = 189\pm12^{o}$ east of north.  The middle panel shows the 178 metal-poor 
  clusters with $\Omega R = 31\pm14$ km s$^{-1}$ and 
  $\Theta_o = 177\pm22^{o}$ east of north, and the bottom panel
  shows the 158 metal-rich clusters with $\Omega R = 47\pm15$ km
  s$^{-1}$ and $\Theta_o = 202\pm15^{o}$ east of north.  The squares represent the
  weighted velocities in $72^{o}$ bins.} 
\label{fig:kin_plot}
\end{center}
\end{figure}

The GCs were assigned individual weights in the sums that combine in
quadrature the individual observational uncertainty, $\varepsilon_v$,  
in $v_r$ and the random velocity
component, $\varepsilon_{random}$, of the GCS.  The dominance of the latter is evident by the large
dispersion in the GC velocities in the kinematic
fitting (see Figure~\ref{fig:kin_plot}).  In other words, the clusters
have individual weights, $\omega_i = (\varepsilon_v^2 +
\varepsilon_{random}^2)^{-1}$; the main purpose of this is to assign a
bit more importance to the clusters with more securely measured
velocities.  This random velocity term dominates in nearly
every case, leaving the GCs with very similar base
weights in the kinematic fitting.  

The three kinematic parameters - rotation amplitude, rotation
axis, and velocity dispersion - are determined with three
different binning methods.  The first involves binning the GCs in radially projected circular annuli from the center of
NGC 5128.   The chosen bins keep a minimum of 15 clusters in each,
ranging as high as 124 clusters.  The bins are 0-5, 5-10, 10-15,
15-25, and 25-50 kpc.  Also, we include 0-50 kpc to determine the
overall kinematics of the system.  

The second method adopts bins with equal numbers of clusters.  The entire population of
clusters had nine bins of 38 clusters each, the metal-poor clusters
had nine bins of 20 clusters, and the metal-rich clusters had eight bins of 20 clusters.
The base weighting is applied to the clusters in both the first and
second binning methods.

The third method uses an
exponential weighting function, outlined in \cite{bergond06}, to
generate a smoothed profile.  This
method determines each kinematic parameter at the radial position, $R$, of
every GC in the entire sample by exponentially
weighting all other GCs surrounding that
position based on their radial separation, $R - R_i$, following

\begin{equation}
\label{eqn:exp_wei}
w_i(R) = \frac{1}{\sigma_R} exp[\frac{-(R - R_i)^2}{2 \sigma_{R}^{2}}].
\end{equation}

In Equation~\ref{eqn:exp_wei}, $w_i$ is the determined weight on each
GC in the sample, and  
$\sigma_R$ is the half-width of the window size.  For this study,
$\sigma_R$ is incrementally varied in a linear fashion for the total sample from $\sigma_R
= 1.0$ kpc at the radius of the innermost
GC in the sample out to $\sigma_R = 4.5$ kpc at the radius of
the outermost GC, where the population is lowest.  The metal-poor population was
given a half-width window of $\sigma_R = 1.0-6.5$ kpc, and the metal-rich
population was given a half-width window of $\sigma_R = 2-5.3$ kpc, again from the
innermost to outermost cluster.   The progressive radial increase in
$\sigma_R$ ensured that each point $R$ had roughly equal total weights.

\subsubsection{Rotation Amplitude of the Globular Cluster System}
\label{sec:rotamp}

The kinematic parameters were determined for the entire sample of 340 GCs, as well as the subpopulations of 178 metal-poor
and 158 metal-rich GCs (four clusters have unknown metallicity). 
The kinematic results for the entire population of GCs
are shown in Table~\ref{tab:all_GC}, reproduced almost in full 
from \cite{w06}, while the results for the metal-poor
and metal-rich clusters are shown in Tables~\ref{tab:MP_GC}
\&~\ref{tab:MR_GC}, respectively.  The
columns give the radial bin, the mean projected radius in the
bin, the radius of the outermost cluster, the number of clusters in the bin, the rotation amplitude, the
rotation axis, and the velocity dispersion, with associated
uncertainties.  These are followed by the mass correction, the pressure-supported
mass, the rotationally supported mass, and the total mass in units of
solar mass (see \S~\ref{sec:dyn} for the mass discussion).  
The results for the alternate two methods, using an equal
number of GCs per bin and the exponentially weighted
GCs, are not shown in tabular form but are included in all of the figures.  

Figure~\ref{fig:kin_plot} shows the sine fit of Eqn.~\ref{eqn:kin} for
the total population and for the metal-poor and metal-rich subpopulations. 
All three populations show rotation about a similar axis.
As discussed in \S~\ref{sec:velfield}, the metal-poor population 
has more members with low velocities ($V_r \leq 300$ km s$^{-1}$) than
the metal-rich population, suggesting possible contamination of
Milky Way foreground stars in the sample.

\begin{figure}
\plotone{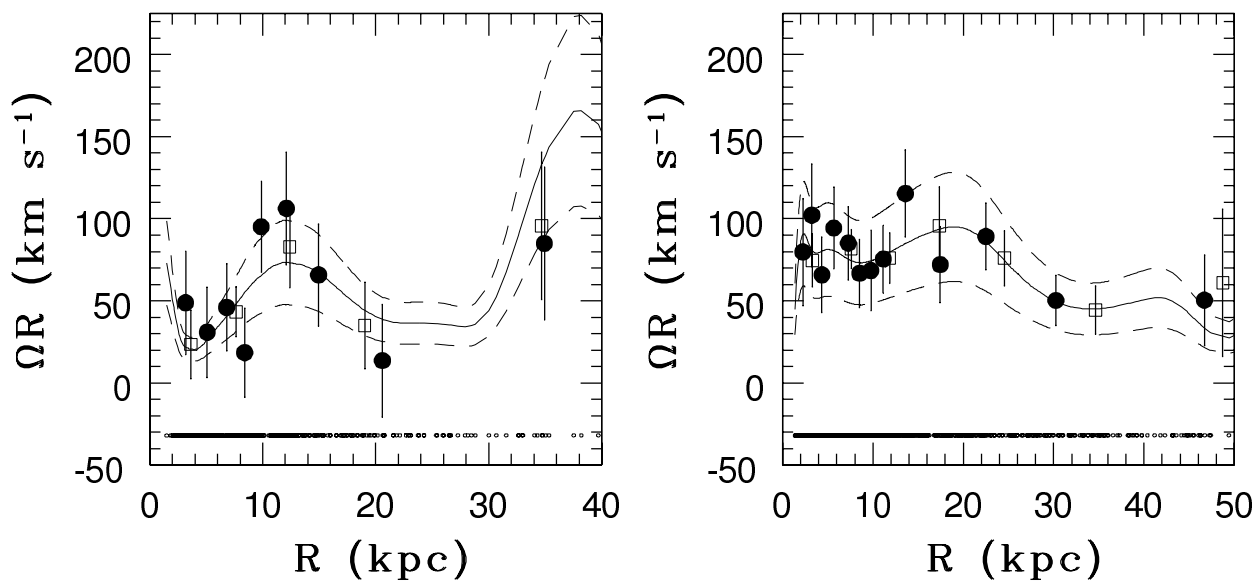}
\caption{Rotation amplitude as a function of projected galactocentric radius
  from the center of NGC 5128, including all known GCs ({\it left})
  and the 780 PNe ({\it right}).
  The data are binned in radial bins ({\it squares}) of 0-5, 5-10,
  10-15 15-25, and 25-50 kpc for the GCs or 0-5, 5-10, 10-15, 15-20,
  20-30, 30-40, and 40-80 kpc for the PNe, in equal number of 38 GCs
  ({\it filled circles}) or 60 PNe, and exponentially weighted GCs or PNe
  with varying bin width ({\it solid line}) and 35\% uncertainties of  
  the weighted data ({\it dashed lines}).  The radial distribution of
  data is shown at the bottom ({\it open circles}).} 
\label{fig:rotamp_final}
\end{figure}

\begin{figure}
\plotone{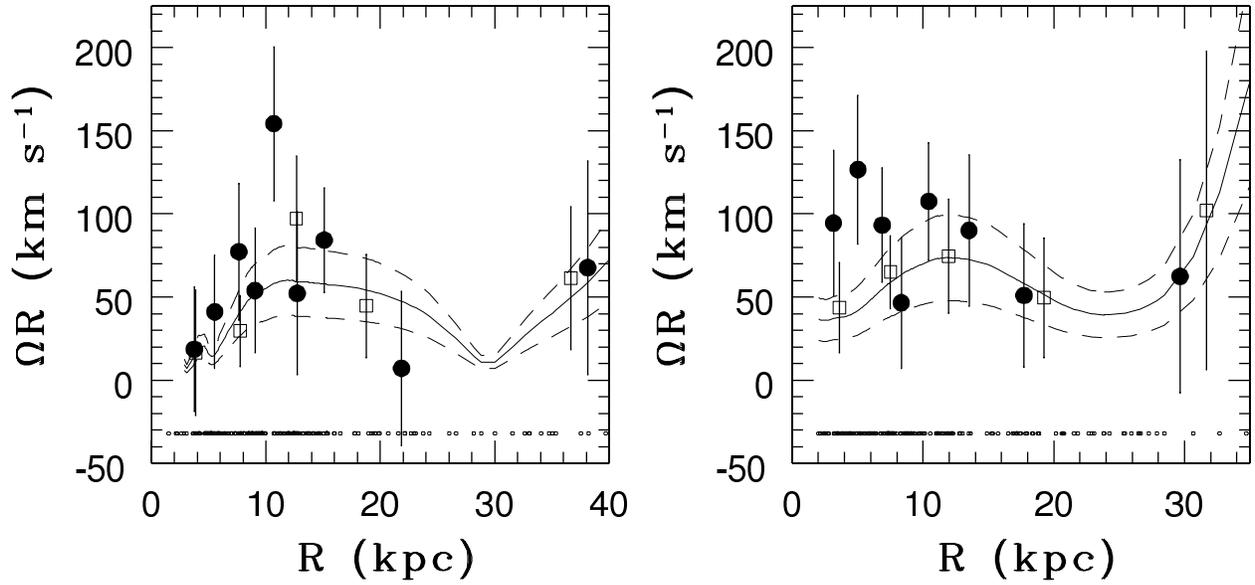}
\caption{Same as Fig.~\ref{fig:rotamp_final}, but for the metal-poor
  subpopulation of GCs ({\it left}) and the metal-rich
  subpopulation of GCs ({\it right}).  The equal numbered bins ({\it
  filled circles}) consist of 20 metal-poor or metal-rich GCs.} 
\label{fig:rotamp_metal}
\end{figure}

Figures~\ref{fig:rotamp_final} \& \ref{fig:rotamp_metal} 
show the rotation amplitude results for the entire population and for the
metal-poor and metal-rich subpopulations, respectively.
The three kinematic methods, described in Section~\ref{sec:math}, 
appear to agree relatively well for all three populations of clusters.  While there appears
to be no extreme difference in rotation amplitude between the cluster
populations, the metal-poor subpopulation of clusters has
lower rotation in the inner 5 kpc of NGC 5128 than the
metal-rich subpopulation.  The weighted average of the 0-5 kpc 
radial bin and the
innermost equal-numbered bin, shows that the entire population 
has a rotation amplitude of $\Omega R = 31\pm17$ km
s$^{-1}$, while the metal-poor population has $\Omega R = 17\pm26$ km
s$^{-1}$ and the metal-rich population has $\Omega R = 57\pm22$ km
s$^{-1}$.  \cite{pff04II} show in their study that the metal-poor 
population has very little rotation
in the central regions, completely consistent with our findings.  
The rotation amplitude does not appear to differ
between the two populations outside of 5 kpc.  

\subsubsection{Rotation Axis of the Globular Cluster System}
\label{sec:rotaxis}

\begin{figure}
\plotone{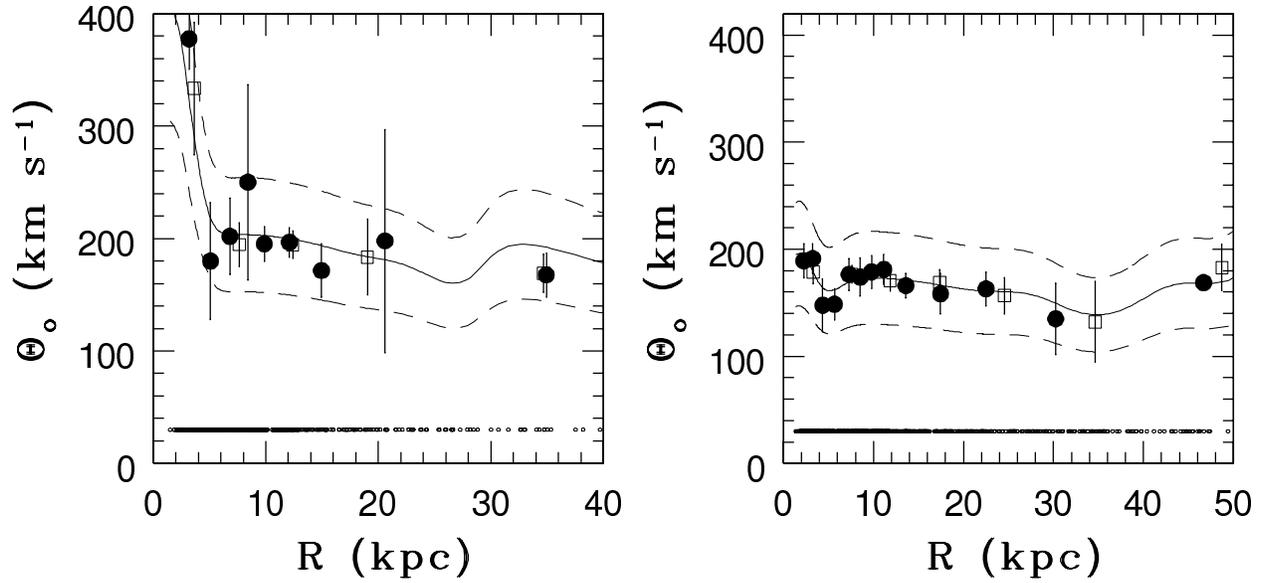}
\caption{Rotation axis as a function of projected galactocentric radius from
the center of NGC 5128, including all known GCs ({\it left}) and
the 780 PNe ({\it right}).  The
symbols are the same as in Fig.~\ref{fig:rotamp_final}, but the uncertainty of  
  the weighted data ({\it dashed lines}) is 25\%.} 
\label{fig:rotaxis_final}
\end{figure}

\begin{figure}
\plotone{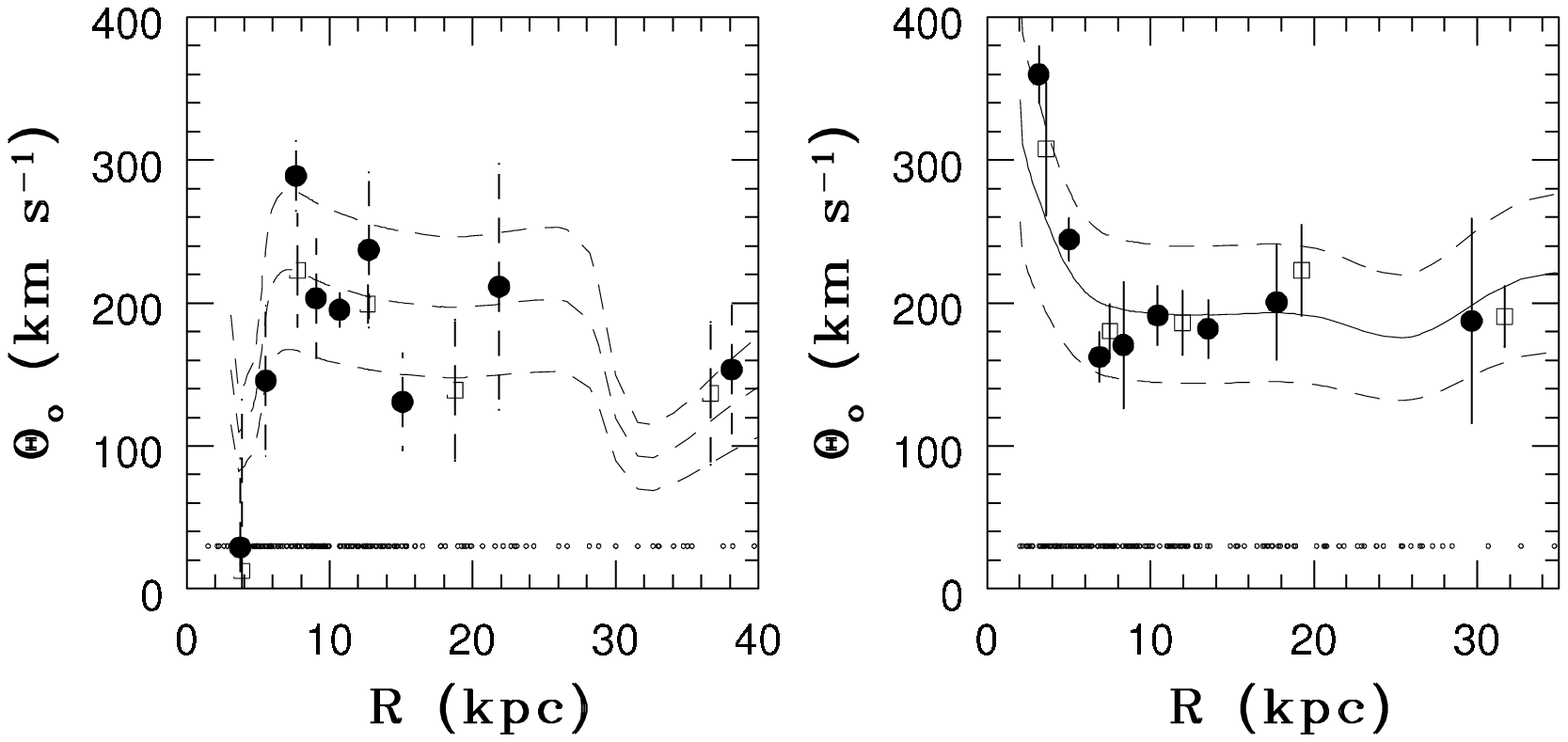}
\caption{Same as Fig.~\ref{fig:rotaxis_final}, but for the metal-poor
  ({\it left}) and metal-rich ({\it right})
  subpopulations of GCs.} 
\label{fig:rotaxis_metal}
\end{figure}

The results of the rotation axis solutions are shown in 
Figures~\ref{fig:rotaxis_final} \& \ref{fig:rotaxis_metal},
again for the entire population and for the
metal-poor and metal-rich subpopulations.
The solution for $\Theta_0$ agrees well
for all three kinematic methods and all subgroups.  
The inner 5 kpc region has a different rotation axis than the 
outer regions, demonstrated clearly in all three binning methods.
The innermost bin yields weighted averages of $\Theta_o = 369\pm24 ^o$, $\Theta_o
= 25\pm55 ^o$, and $\Theta_o = 352\pm18 ^o$, all of which are equal
within their uncertainties.  Beyond 5 kpc, the rotation axes for all three populations are in
even closer agreement, with averages of $\Theta_o = 189\pm6 ^o$,
$\Theta_o = 199\pm7 ^o$, and $\Theta_o = 196\pm7 ^o$ 
for the entire population, the metal-poor subpopulation, and the metal-rich subpopulation,
respectively.  The position angle of the photometric major axis of NGC
5128 is $\Theta = 35^o$ and $215^o$ east of north and the photometric minor
axis is $\Theta = 119^o$ and $299^o$ east of north \citep{dufour79}.  It
appears the GCS is rotating about an axis similar to the photometric
major axis for the full extent of the galaxy, with a possible axial
twist or counterrotation within 5 kpc.

\subsubsection{Velocity Dispersion of the Globular Cluster System}
\label{sec:veldisp}

\begin{figure}
\plotone{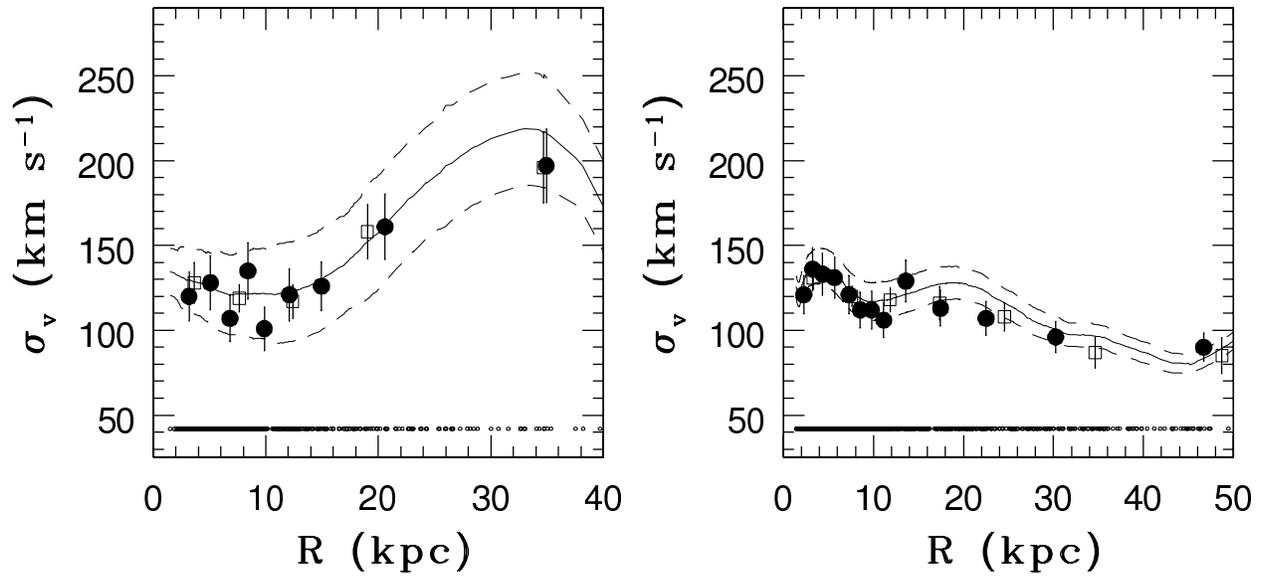}
\caption{Velocity dispersion as a function of projected
  galactocentric radius from the center of NGC 5128 for all known GCs
  ({\it left}) and the 780 PNe ({\it right}).  The
symbols are the same as in Fig.~\ref{fig:rotaxis_final}, but the uncertainties of  
  the weighted data ({\it dashed lines}) are determined by Eqn~\ref{eqn:veldisp}.} 
\label{fig:veldisp_final}
\end{figure}

\begin{figure}
\plotone{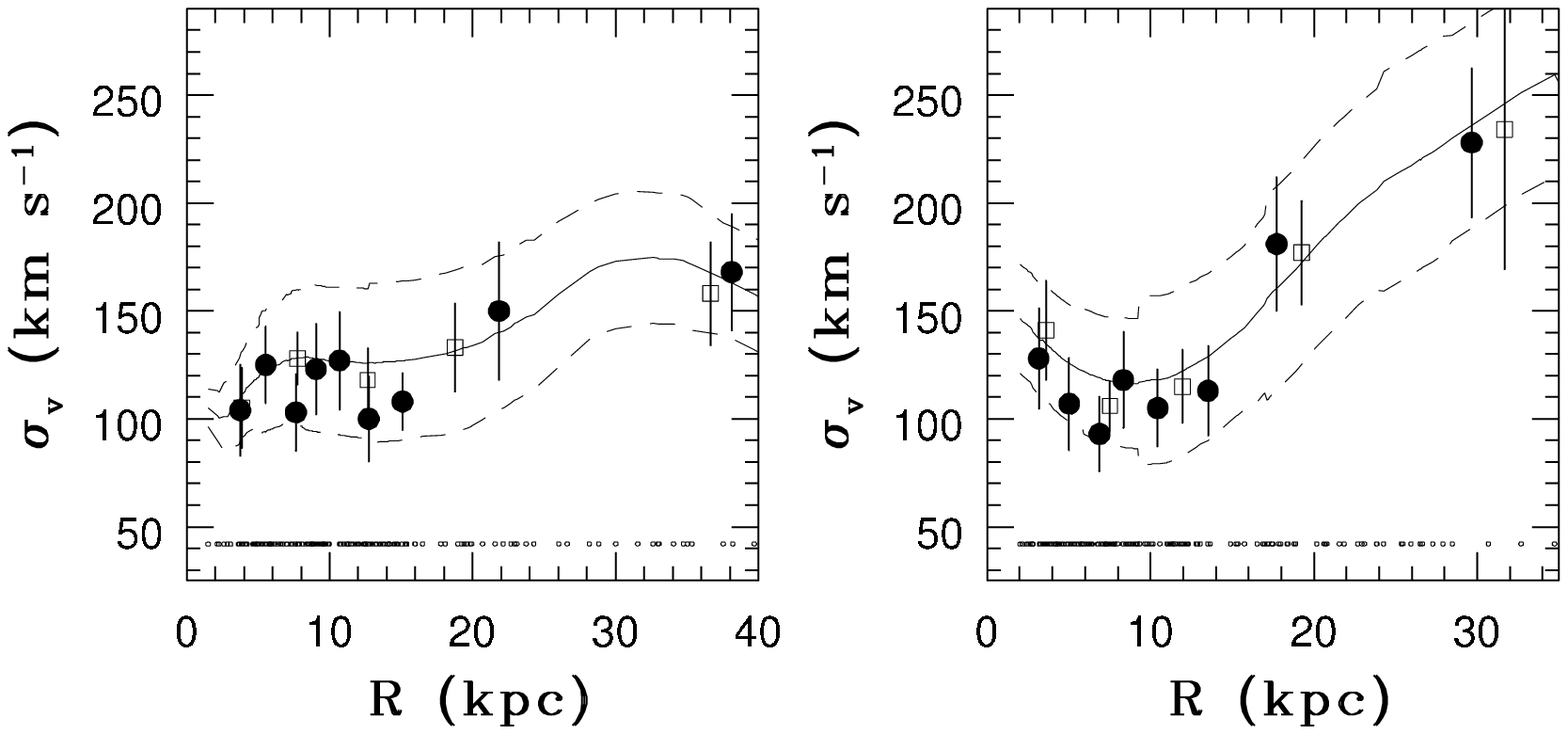}
\caption{Same as Fig.~\ref{fig:veldisp_final}, but for the
  metal-poor ({\it left}) and metal-rich ({\it right})
  subpopulations of GCs.} 
\label{fig:veldisp_metal}
\end{figure}

Figures~\ref{fig:veldisp_final} \& \ref{fig:veldisp_metal}
show the velocity dispersion for the entire population and for the
metal-poor and metal-rich subpopulations. 
Our results for $\sigma_v$ show no
significant differences between the metallicity subpopulations.  All three show a relatively flat
velocity dispersion ($\sigma_v = 119\pm4$, $\sigma_v = 117\pm6$, and $\sigma_v = 111\pm6$ km s$^{-1}$ within 15 kpc
of the center of NGC 5128 for the entire population and for the
metal-poor and metal-rich subpopulations, respectively).  These results
match the previous study of NGC 5128 by \cite{pff04II},
whose determined velocity dispersion for the GCs within
20 kpc ranged between 75 and 150 km s$^{-1}$.
At a larger radius, we find that $\sigma_v$
then slowly increases to $\sigma_v > 150$ km s$^{-1}$  
towards the outer regions of the halo for all populations.  
The velocity dispersion of the metal-rich GCs,
interestingly, appears higher than that of the metal-poor GCs in the outer regions (although still
consistent within the determined uncertainties).  In most previous
studies, the velocity dispersion of the metal-poor GCs usually appears higher
than that of the metal-rich GCs, if there is a notable velocity
dispersion difference between the
subpopulations \cite[see the studies of][as examples]{cote03,richtler04}.

\begin{figure}
\plotone{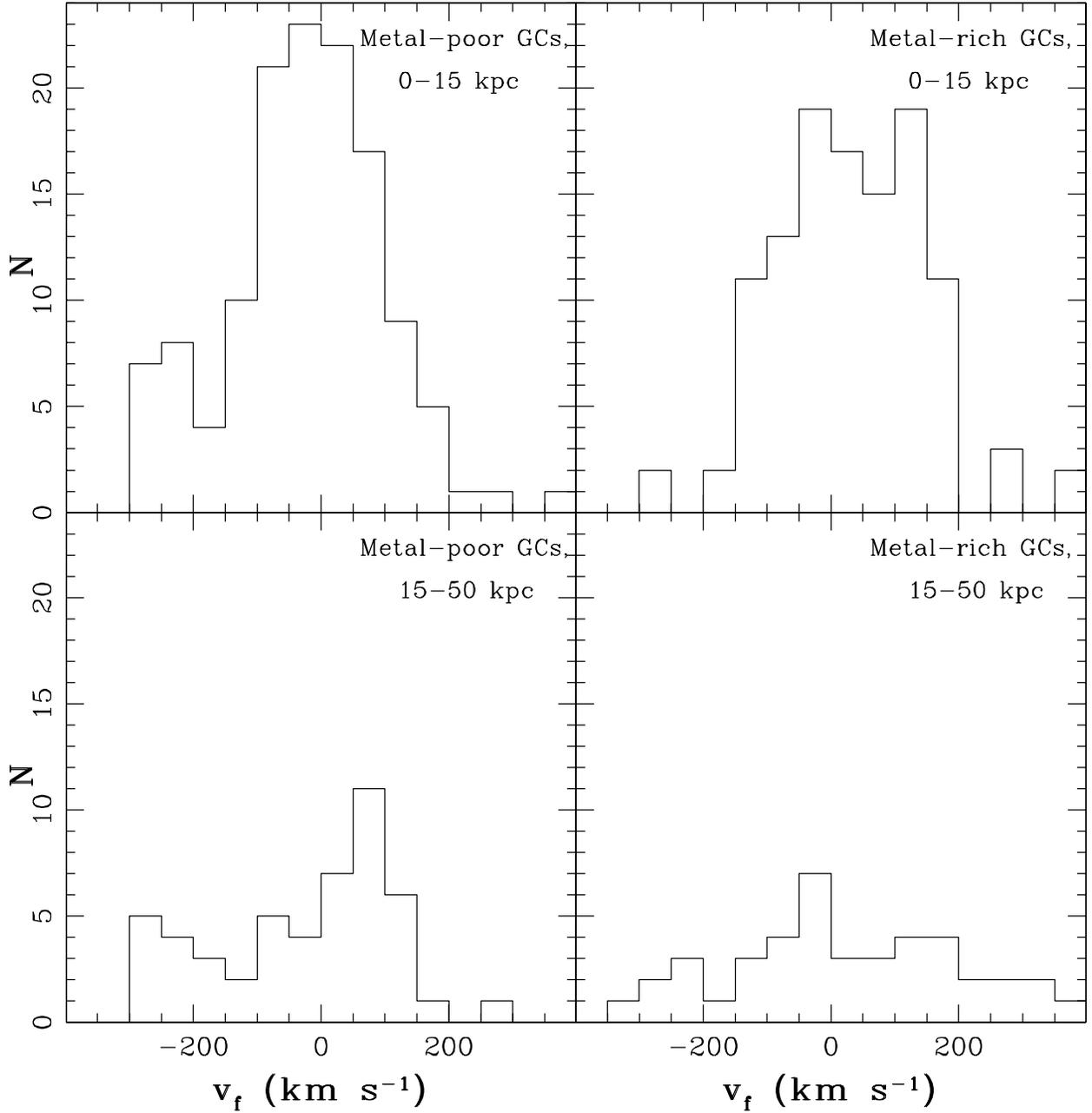}
\caption{Velocity histograms for the GCs, subdivided by
  both metallicity and radius, binned in 50 km s$^{-1}$ intervals.  Here $v_f$ is the residual
  velocity after subtraction of the systemic velocity of 541 km
  s$^{-1}$ and the overall rotation component for either the
  metal-poor or metal-rich subpopulation, determined from
  Eqn.~\ref{eqn:kin}.   The histograms for the metal-poor clusters
  ({\it left panels}) include 129 clusters between 0 and 15 kpc and 49
  clusters between 15 and 50 kpc, while the histograms for the metal-rich clusters 
  ({\it right panels}) include 114 clusters between 0 and 15 kpc and 44 clusters
  between 15 and 50 kpc.}
\label{fig:vf_histo}
\end{figure}

To explore the cause of the distinct rise past 15 kpc a bit further, we have plotted
the actual velocity histograms in
Figure~\ref{fig:vf_histo} for the metal-poor and metal-rich subgroups,
subdivided further into inner ($R < 15$ kpc) and outer ($R > 15$ kpc) regions.
In the inner 15 kpc, both samples show histograms strongly peaked near
$v_f = 0$ and with at least roughly Gaussian falloff to both high
and low velocities.  By contrast, the histograms for the outer regions
($15-50$ kpc) are noticeably flatter, so that the clusters with larger velocity residuals have
relatively more importance to the formal value of $\sigma_v$.  Nominally,
the flatter shape of the velocity distribution would mean that the outer-halo
clusters display anisotropy in the direction of a bias towards more
circular orbits.  However, such a conclusion would be premature at this
point for two reasons.  First, the sample size in the outer regions is still
too small to lead to high significance, and a direct comparison between
the inner and outer histograms (through a Kolmogorov-Smirnov test)
does not show a statistically significant difference between them larger
than the 70\% level.  Second, the outer samples may still be spatially
biased in favor of objects along the major axis of the halo, as discussed
above, and this bias sets in strongly for $R > 12$ kpc 
(see Fig.~\ref{fig:gc_thetar}),
very near where we have set the radial divisions in this Figure.  This
type of velocity distribution can also arise from the accretions of
satellite galaxies with their own small numbers of GCs \citep{bekki03}.
We will need to have a larger sample of the outer-halo clusters, and one
in which these potential sample biases have been removed, before we can
draw any firmer conclusions.  However, it needs to be explicitly
stated that the outermost point in the kinematic plots for the GCs
representing $25-50$ kpc suffers from very high spatial biases and 
low number statistics ($< 40$ GCs) and covers a large radial interval.
The rise in velocity dispersion could be driven purely by systematic
effects resulting from the radial gradient of the number density of
GCs in this outermost bin \citep{napolitano01}.

\section{Kinematics of the Planetary Nebula System of NGC 5128}
\label{sec:kin_PN}

NGC 5128 has a large number, 780, of identified PNe with
measured radial velocity from the studies of
\cite{peng04} and \cite{hui95}; these PNe are projected out to 90 kpc assuming a
distance to NGC 5128 of 3.9 Mpc.  Since these are
also old objects, it is of obvious interest to compare them with the
GCS.  The PNe also have the advantage
of giving us the best available look at the kinematics of the halo field stars.

The PN kinematic results are listed in
Table~\ref{tab:PN}, with the same columns as Table~\ref{tab:all_GC}.
The results are shown in Figures~\ref{fig:rotamp_final},
\ref{fig:rotaxis_final}, \&~\ref{fig:veldisp_final} for the rotation
amplitude, rotation axis, and velocity dispersion, respectively.  The
spatial distribution of the known PNe is, like the
GCS, biased toward the major axis at large radii
(see Fig.~\ref{fig:position}).  Nevertheless, their kinematics closely resemble the
GCs.

The kinematics of the PN system are very consistent
among all three binning methods.
The rotation amplitude and rotation axis show little radial trend,
while the velocity dispersion appears relatively flat within the first
15 kpc at $\sigma_v = 122\pm7$ km s$^{-1}$ and then slowly {\it decreases}
to $\sigma_v \simeq 85$ km s$^{-1}$ at large galactocentric radius.
\cite{pff04I} show that the velocity dispersion of the PNe
drops from a central value of 140 to 75 km s$^{-1}$ in the
outer regions of the galaxy, consistent with the findings of this
study.  Their velocity field analysis led to the discovery of a ''zero-velocity curve'' located
between the photometric minor axis, $119\pm5 ^o$ east of north
\citep{dufour79}, and the north-south
direction, for the innermost region of the galaxy.  
Just beyond 5 kpc, the zero-velocity curve turns and follows a
straight line at a $7^o$ angle from the photometric major axis,
$35^o$ east of north \citep{dufour79}
\citep[see Figure 7 of][]{pff04II}. 

This study does not show a strong change in rotation axis for the
PNe in the innermost regions of the galaxy.
However, it clearly shows in all three GC populations 
a significant change in the rotation axis just beyond 5 kpc from the 
center of the galaxy.  A change in axis of 5 kpc outward ($\sim180 ^o$ for the 
entire population and metal-poor subpopulation and $\sim160 ^o$ 
for the metal-rich subpopulation) has been found, as discussed in
\S~\ref{sec:rotaxis}.  Similarly, \cite{pff04II} 
show from their sample of 215 clusters that
a clear sign of rotation beyond 5 kpc about a misaligned axis appears 
particularly in their metal-rich subpopulation.  The
kinematics of the GCs in this study matches the line of
zero velocity relatively nicely.  Within 5 kpc the rotation axis of the GCS is nearly-parallel to the north-south direction, and beyond 5
kpc the rotation axis is near $200 ^o$ east of north, which is only
$\sim10^o$ from the zero-velocity curve. 

However, the velocity field of NGC 5128 is complex and not entirely
captured by these approximate solutions.  The
two-dimensional velocity field shown in \cite{peng04} (see their
Figure~7), shows that the photometric major axis (which happens to be
very close to our maximum rotation as discussed above) is 
only $7^o$ from the line of zero velocity.  This could lead to a 
very asymmetric velocity profile that may not be well fit by 
the sine curve described in Equation~\ref{eqn:kin}.  Biased
kinematics, especially the rotation axis,
may develop from the sine fit that could lead to a higher estimated
velocity dispersion.  

\cite{hui95} similarly studied the kinematics of the PN
system in NGC 5128 with a sample of 433 PNe.  They obtain a rotation axis
of $344\pm10^o$.  Our result of $170\pm5^o$ east of north is
consistent with their findings on comparing their sine curve fit of
their PN data in their Figure~11 to our corresponding fit shown in
Fig.~\ref{fig:kin_plot} for the GCS, which shares a similar axis to
our PN sample (note that in their study, $\phi = 0 ^o$ corresponds to our
$\Theta = 305^o$ east of north).  Our fits both correspond to a positive
rotation amplitude for a rotation axis near $170^o$ east of north and a
negative rotation axis near $350^o$ east of north.  Therefore, the rotation
axis quoted in \cite{hui95} of $344\pm10^o$ corresponds to a {\it negative} rotation
amplitude of approximately $70-75$ km s$^{-1}$ (taken from their
Figure~11), nicely matching our
result of a positive $76\pm6$ km s$^{-1}$ about an axis of $170\pm5^o$
east of north.

\section{Dynamics of NGC 5128}
\label{sec:dyn}

Both GCs \citep[][among
  others]{cote01,larsen02,evans03,cote03,beasley04,pff04II} and 
PNe \citep[][among others]{ciardullo93,hui95,arnaboldi98,peng04}
can be used to estimate the total dynamical mass of their host
galaxies.  A variety of tools are in use including derived mass models, the virial mass estimator \citep{bahcall81},
the projected mass estimator \citep{heisler85}, and the tracer mass
estimator \citep{evans03}. 

NGC 5128 does not have a large X-ray halo \citep[detected by][the latter
reporting a measurement of log $L_x = 40.10$ erg s$^{-1}$]{kraft03,osullivan01}, such as is evident in other giant
ellipticals such as M87 \citep{cote01} or NGC 4649 \citep{bridges06}.  
Thus it is difficult to model the dark matter
profile of NGC 5128 with {\it a priori} constraints.  Without such a mass model, we turn to the
tracer mass estimator for the dynamical mass determination.  The
tracer mass estimator has the distinct advantage over the virial and projected
mass estimators that the tracer population does not have to follow the
dark matter density in the galaxy - an extremely useful feature for
stellar subsystems such as GCs and PNe
that might, in principle, have significantly different radial
distributions (see \cite{evans03} for extensive discussion).  Below,
we determine the mass of NGC 5128 using the tracer populations of
GCs and PNe (our mass estimates do not
include stellar kinematics in the inner regions).

\subsection{Mass Determination}
\label{sec:mass}

The mass contributed by the random internal motion of the galaxy
(pressure-supported mass) is
determined from the tracer mass estimator as
\begin{equation}
\label{eqn:tme}
M_{p} = \frac{C}{GN} \sum_{i}(v_{f_i} - v_{sys})^2R_i
\end{equation}
where $N$ is the number of objects in the sample and $v_{f_i}$ is the
radial velocity of the tracer object {\it with the rotation
component removed}.  For an isotropic population of tracer objects, assumed in this study, the value of $C$ is
\begin{equation}
\label{eqn:C}
C = \frac{4(\alpha + \gamma)(4 - \alpha -\gamma)(1-(\frac{r_{in}}{r_{out}})^{(3-\gamma)})}{\pi(3-\gamma)(1-(\frac{r_{in}}{r_{out}})^{(4-\alpha-\gamma)})}
\end{equation}
where $r_{in}$ and $r_{out}$ are the three-dimensional radii corresponding
to the two-dimensional projected radii of the
innermost,  $R_{in}$, and outermost, $R_{out}$, tracers in the
sample. 
The parameter $\alpha$ is set to zero for an isothermal 
halo potential in which the system has a flat
rotation curve at large distances.  Finally, $\gamma$ is
the slope of the volume density distribution, which goes as $r^{-\gamma}$,
and is found by determining the surface density slope of the sample and
deprojecting the slope to three-dimensions.  
The tracer mass estimator uses a sample of tracer objects defined between $r_{in}$ and
$r_{out}$, yet it is important to emphasize that it determines the 
{\it total} enclosed mass within $r_{out}$. 

There is also a contribution to the total mass by the rotational
component, as determined in \S~\ref{sec:rotamp} for the GCs and \S~\ref{sec:kin_PN} for the PNe.  
This mass component is determined from the rotational
component of the Jeans equation,
\begin{equation}
\label{eqn:rje}
M_{r} = \frac{R_{out}v^{2}_{max}}{G}
\end{equation}
where $R_{out}$ is the outermost tracer projected radius in the sample and
$v_{max}$ is the rotation amplitude.
Therefore, the total mass of NGC 5128, $M_t$, is determined by the addition of the
mass components supported by rotation, $M_r$, and random internal
motion, $M_p$,
\begin{equation}
\label{eqn:total_mass}
M_{t} = M_{p} + M_{r}.
\end{equation}

In the determination of the pressure-supported mass, one must estimate
values for $r_{in}$ and $r_{out}$ knowing $R_{in}$ and $R_{out}$.  \cite{evans03} suggest that $r_{in}\simeq
R_{in}$ and $r_{out}\simeq R_{out}$ for distributions taken over a
wide angle.  However, in this study the inner and outer radii
of the chosen bins are at intermediate radial values within the
distribution.  Their assumption would therefore lead to an
underestimate of the determined mass, since the true $r_{out}$ can be
quite a bit larger than the projected $R_{out}$.  To correct for this contributed
uncertainty, distributions of sample tracer populations were generated
through Monte Carlo simulations.  In the simulations, 340 GCs
were randomly placed in a spherically symmetric system extending out
to 50 kpc with an $r^{-2}$ projected density, while 780 PNe were placed in
the same environment extending out to 90 kpc.
From the generated distributions, the value of $C$ in Eqn.~\ref{eqn:C}
was determined for both the real and projected positions of the tracer
populations in each designated radial bin.  This correction factor, listed in
Tables~\ref{tab:MP_GC}-\ref{tab:PN} as $M_{corr}$, multiplies the
pressure-supported mass from Eqn.~\ref{eqn:tme}.  The same correction
was applied to the full GC sample and the corresponding
$M_{corr}$ values are listed in Table~1 of \cite{w06}. 
These values are generally small, but in the worst case they triple $M_p$.
 
\subsection{Surface Density Profiles}
\label{sec:surden}

\begin{figure}
\plotone{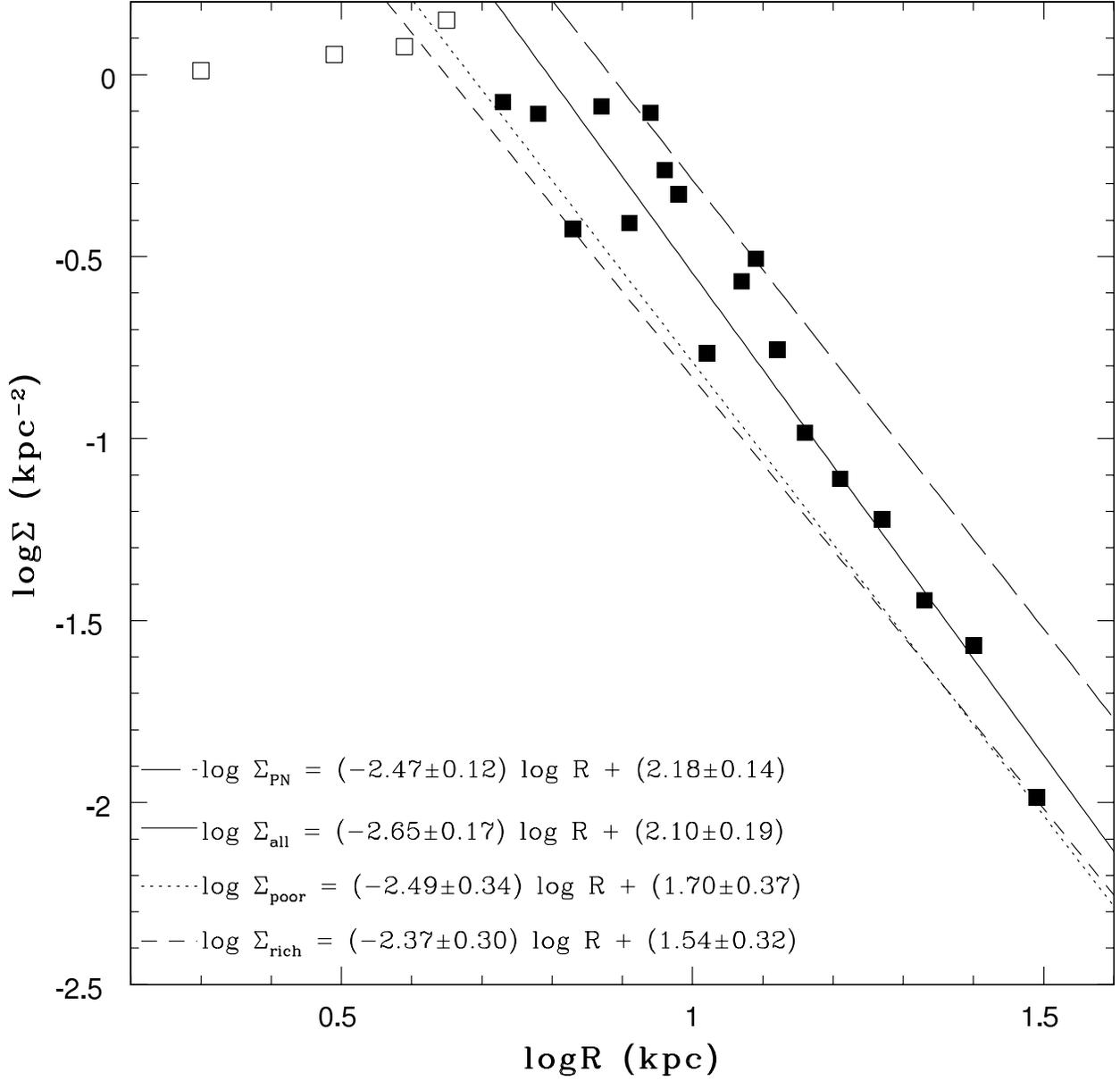}
\caption{Surface density of known GCs fit with a
  powerlaw for the entire population ({\it solid line}), the
  metal-poor subpopulation ({\it dotted line}), and the metal-rich subpopulation ({\it short-dashed line})
  in NGC 5128 with the best-fit linear coefficients shown.  
  The radial distribution of the entire
  GC population is shown as filled squares, yet
  clusters with projected radii $< 5$ kpc are shown as open
  squares and are not included in the powerlaw fits.  Also 
  overplotted is the surface density of the PN 
  system ({\it long-dashed line}) fit from all PNe beyond 5 kpc with the 
  best-fit linear coefficients also shown.} 
\label{fig:gc_surden}
\end{figure}

In Eqn.~\ref{eqn:C}, the value of $\gamma$ is determined for the
tracer populations by deprojecting the slope of the surface density
profile to three-dimensions.  Figure~\ref{fig:gc_surden} shows the
surface density profiles for the entire, metal-poor, and metal-rich GC
populations, along with the PN profile.   The populations
were binned, following \cite{maiz05}, into circular annuli of equal numbers of
objects, providing the same statistical weight to each bin (although spatial
biases may still affect the GC population in the outer regions along the major axis;
see Fig.~\ref{fig:gc_thetar}).  In the inner 5 kpc of all tracer
populations, incompleteness due to the obscuration
of the dust lane is evident by the flattening of the surface density
profile.  The innermost objects were, therefore, excluded from the
surface density profile fittings.  Outside of 5 kpc, the surface
densities fit well to power laws, leading to $\gamma = 3.65\pm0.17$,
$3.49\pm0.34$, $3.37\pm0.30$, and $3.47\pm0.12$ for the entire
GC population, the metal-poor and metal-rich
subpopulations of GCs, and the PNe in NGC
5128, respectively.  These are all very similar within their uncertainties.

\subsection{Mass Results}
\label{sec:mass_results}

The similar kinematics we find between the metal-poor and
metal-rich subpopulations of GCs in this study strongly justifies
the combining of the two populations for the mass determination
performed in \cite{w06}.
The GC population provides a {\it total} mass estimate of
$(1.3\pm0.5) \times 10^{12}$  $M_{\odot}$ from 340 clusters out to a
projected radius of 50 kpc.  Removing the GCs in our sample with $v_r
\leq 300$ km s$^{-1}$, which will remove all possible contamination
from foreground stars, discussed in \S~\ref{sec:velfield}, leads
to a total mass of $(1.0\pm0.4)  \times 10^{12}$ $M_{\odot}$.  This mass
agrees nicely with our mass determined from our entire GC sample.   The PN
population provides a total mass of $(1.0\pm0.2) \times 10^{12}$ $M_{\odot}$ from 780 PNe out to 90 kpc in projected radius,
agreeing with the GC value within the uncertainty.  

\begin{figure}
\plotone{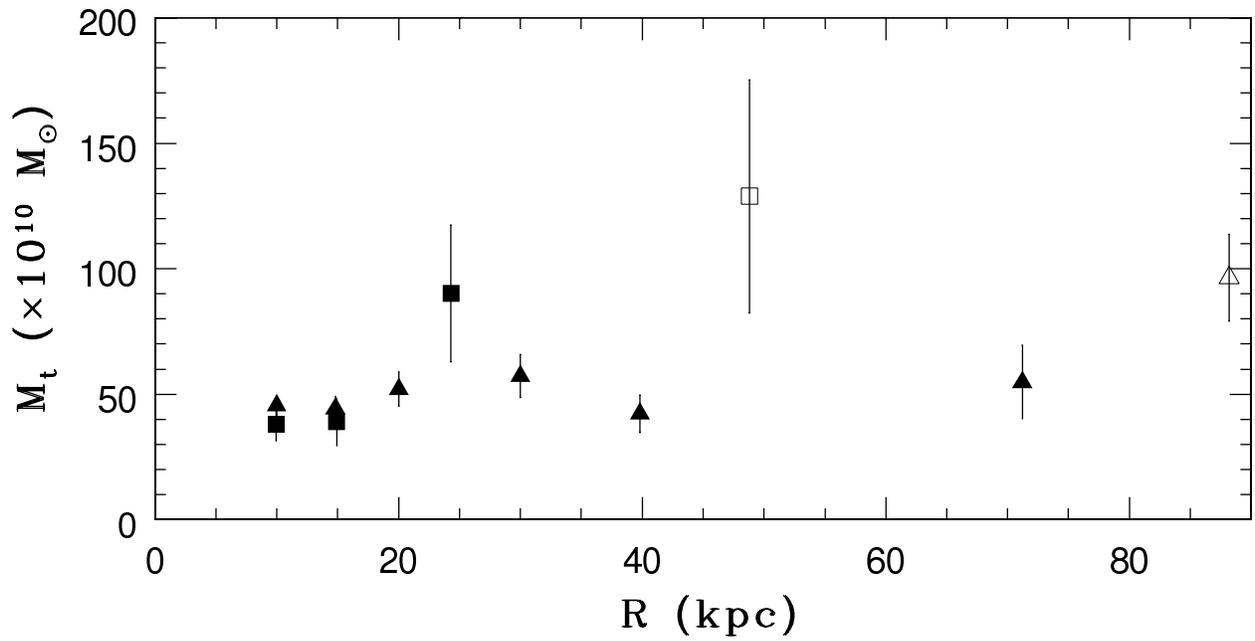}
\caption{Total mass profile for the radially binned
  estimates of the GC
  system ({\it filled squares}) from \cite{w06} and the PNe ({\it filled
  triangles}).  The GC total mass from 0 to 50 kpc,
  shown as the open square, is $1.3\pm0.5 \times 10^{12}
  M_{\odot}$, and the PN total mass from 0 to 90 kpc,
  shown as the open triangle, is $1.0\pm0.2 \times 10^{12}
  M_{\odot}$, agreeing within the uncertainties.}
\label{fig:mass}
\end{figure}

We are also able to generate a mass profile of NGC 5128 from the total
GC population and the PNe, shown in Figure~\ref{fig:mass}.  
The tracer mass
estimator determines the {\it total} enclosed
mass for NGC 5128 within the outermost radius of a given tracer
sample.  It calculates this total mass using a sample of objects defined within the radial range
defined by the sample's inner and outermost radii.  
It is therefore possible to use a unique set of tracer objects,
denoted by the radial bin range, listed in the first column of 
Tables~\ref{tab:all_GC}-\ref{tab:PN}, to determine a mass profile from independent mass
estimates.  The independent binning, leads to sample sizes in the mass
determination, in some cases generating higher uncertainties in the total enclosed mass.  
The most certain mass is the one determined from the full
sample of tracers.  

In the mass determinations above, we have implicity assumed isotropy
for the velocity distributions.  But the possibility exists that the
PNe (for example) might have radial anisotropy which would produce
their gradually falling $\sigma_v(R)$ curve.  Replacing Equation~\ref{eqn:C} in the tracer mass estimator by
\begin{equation}
\label{eqn:C_aniso}
 C = \frac{16(\alpha + \gamma - 2\beta)(4 - \alpha
   -\gamma)(1-(\frac{r_{in}}{r_{out}})^{(3-\gamma)})}{\pi(4 - 3\beta)(3-\gamma)(1-(\frac{r_{in}}{r_{out}})^{(4-\alpha-\gamma)})}
\end{equation}
which includes the anisotropy parameter, $\beta$, from \cite{evans03},
we find that the mass estimate from the PNe can be forced to agree 
with the mass estimate from the GCS for a nominal $\beta = 0.8$.  For
perfect isotropy, $\beta = 0$.  This 
would mean roughly 2:1 radial anisotropy for the PNe in the
outer halo.  However, we find that any 
$\beta$ in the wide range
of $-10 \leq \beta \leq 1$ would still keep the two methods in agreement within
their internal uncertainties, so we are not yet in a position to
tightly constrain any anisotropy.  It is possible that the GCs may
also have anisotropy; it may therefore be too simplistic to find a
range of $\beta$ for the PNe for which the masses of the PNe and GCs
agree.  However, the GCs are likelier to be nearly isotropic than the
PNe; the GCs are older, ''hotter'' subsystems of the halo.  In other
studies, the isotropy of the GCS orbits has also been shown to be a good
assumption from mass profiles of elliptical galaxies using X-ray
observations \citep[][among others]{cote01, cote03, bridges06}.

Both mass estimates can be compared to previous studies.  First, we
note the
total mass determined from the PN data with that of
\cite{peng04}.  While the rotationally supported mass was 
determined here with different values of the mean rotational velocity,
they calculated the pressure supported mass using the identical tracer
mass estimator technique with exactly the same PN
population.  The total mass estimate given by \cite{peng04} is
$(5.3\pm0.5) \times 10^{11}$ $M_{\odot}$.  Subtracting their rotationally
supported mass leaves a pressure supported mass of $\sim 3.4 \times
10^{11}$ $M_{\odot}$, quite different from our 
$(8.46\pm1.72) \times 10^{11}$ $M_{\odot}$.  Recalculating our pressure
supported mass estimate with $\gamma = 2.54$, which was used in \cite{peng04}, we are able
to reproduce their mass estimate within the uncertainty.  
The values of $\gamma$ differ between the two studies simply because 
the $\gamma$ used in \cite{peng04} was the inverse of the surface density slope instead of
the inverse of the volume density slope.  Using the correct value of $\gamma = 3.54$, their
pressure supported mass estimate would increase to $8.7 \times
10^{11} M_{\odot}$, matching the mass found in this study.  

Second, we compare our total mass determined using the GC 
population with that from \cite{pff04II}.  Using 215 GCs out to 40 kpc, they found a
pressuresupported mass of $(3.4\pm0.8) \times 10^{11}$ $M_{\odot}$,
again much different from our pressure supported mass of $(1.26\pm0.47) \times 10^{12}$ $M_{\odot}$
using 340 clusters out to 50 kpc.  The large
difference can again be attributed to their using $\gamma = 2.72$ instead of
deprojecting their surface density slope to $\gamma = 3.72$.  
Using the correct value of $\gamma$, we find a pressure-supported mass of $7.5 \times 10^{11}$ $M_{\odot}$ using the same 215
clusters they used in their study.  This corrected estimate is closer to the pressure
supported mass determined in our study, but it is not necessarily expected
to agree with our result, as our sample contains $130$ more GCs
and uses a slightly different $\gamma$ that we have independently
redetermined.  

Third, the mass
determined by the H $_I$ shell study of NGC 5128 by
\cite{schiminovich94} found a mass of $2 \times 10^{11}$ $M_{\odot}$
{\it within 15 kpc} assuming a distance of 3.5 Mpc.  With the distance of 3.9
Mpc used in this study, the mass determined in their study would
increase to $2.2 \times 10^{11}$ $M_{\odot}$, which is $30\%$ smaller
than our total mass of $3.89\pm0.94 \times
10^{11}$ $M_{\odot}$ within 15 kpc.  

Lastly, we compare our determined mass to a recent study by
\cite{samurovic06}.  \cite{samurovic06} determined a total mass of NGC 5128 using 
GCs, PNe, and an X-ray data technique.  The galaxy mass determined
from the GC and PN data was obtained using  
the tracer mass estimator and the spherical Jeans equation, as
performed in our study.  However, \cite{samurovic06} used the volume density slopes
determined by \cite{peng04} and \cite{pff04II}, for the PN and GC
data, respectively.  They obtained
mass estimates for NGC 5128 similar to those of \cite{peng04} and
\cite{pff04II}, discussed above, using an identical PN sample and
slightly increased GC sample.  They also
included an X-ray-modelling mass estimate for NGC 5128 from which they
obtained masses of $(7.0\pm0.8) \times 10^{11}$ $M_{\odot}$ out to 
50' and $(11.6\pm1.0) \times 10^{11}$ $M_{\odot}$ out to 80'.  
This mass estimate is similar to our PNe estimate out to the
same radial extent, but the author cautions that it is an
overestimate of the true mass of NGC 5128 resulting from a lack of
hydrostatic equilibrium in the outer region of the galaxy. 

We note here that the mass estimates obtained are higher than those
from \cite{hui95}, and \cite{peng04} derived from the PNe using a
two-component mass model, as well as \cite{samurovic06}, using an X-ray
modelling technique.  This discrepancy is not fully understood
and possibly lies in the assumptions that go into the mass estimators
with a spatially biased sample.
We intend to pursue this issue further with an upcoming larger sample of GCs with
less spatial biases. 

The mass of NGC 5128 that we find appears to be in the range of other giant elliptical
galaxies, such as NGC 1399 \citep[$\sim 2 \times 10^{12}$ $M_{\odot}$ out to 
50 kpc;][]{richtler04}, M49 \citep[$\sim 2 \times 10^{12}$ $M_{\odot}$ 
out to their kinematically studied radius of 35
kpc;][]{cote03}, and M87 \citep[$\sim 9 \times 10^{11}$ $M_{\odot}$
at 20 kpc, the onset of the projected cD envelope;][]{cote01}.  
Clearly, it is legitimate to say that NGC 5128 is the
largest, most massive galaxy in the neighborhood of the Local Group,
and one that can be talked about in the same category as these other
giants that reside in larger clusters.

The sample biases mentioned above in our currently available set of
both GCs and PNe place limitations on how much we can reasonably
interpret the kinematic and dynamic data.  We are currently carrying
out a set of new spectroscopic programs to increase the tracer sample
size and to remove the sample biases, leading to a more complete
analysis of the halo velocity field.

\section{Discussion and Conclusions}
\label{sec:concl}

Angular momentum is an essential quantity for characterizing the sizes,
shapes, and formation of galaxies and is often represented as the 
dimensionless spin parameter,

\begin{equation}
\label{eqn:spin}
\Lambda = \frac{J |E|^{1/2}}{G M^{5/2}}
\end{equation}
where $J$ is the angular momentum, $E$ is the binding energy, and $M$
is the mass of the galaxy.  The spin parameter is representative of a galaxy's angular
momentum compared to the amount of angular momentum needed
for pure rotational support:  the lower the $\Lambda$-value, the less
rotation and rotational support within the galaxy.  
For an elliptical galaxy in
gravitational equilibrium, the spin parameter simplifies to $\Lambda
\sim 0.3 <(\Omega R / \sigma_v)>$ \citep{fall79}, yielding $\Lambda =
0.10$ with $(\Omega R / \sigma_v) = 0.33$ for the entire population
of GCs in NGC 5128.  

Table~\ref{tab:spin} shows the spin parameter for four giant galaxies
with large GCS kinematic studies, M87, M49, NGC 1399, and NGC 5128.  
The table columns give the galaxy name, the
rotation amplitude, the projected velocity dispersion, and the ratio of
the rotation amplitude to the velocity dispersion, followed by the spin
parameter.  These quantities are shown for the metal-poor and
metal-rich populations.  What is clearly evident in
Table~\ref{tab:spin} is the strong galaxy to galaxy differences
between these four galaxies, already hinted at in
\S~\ref{sec:intro}.  Though the sample is still quite small, no
obvious pattern emerges.  There is an indication
of metal-poor and metal-rich GCSs having similar spin parameters
within the same galaxy.  M49 is the only galaxy studied here where this may not be
the case.  Although the metal-poor and metal-rich cluster spin
parameters are consistent within the uncertainties, the metal-rich cluster
spin parameter of M49 is also consistent with zero. 

In the monolithic collapse scenario, \cite{peebles69} describes the angular
momentum within the galaxy as attributed to the tidal torque
transferred from neighbouring proto-galaxies during formation.  In
this scenario, \cite{efstathiou79} found that a spin parameter of
$\Lambda = 0.06$ for elliptical systems is expected from simulations
of the collapse of an isolated protogalactic cloud.  But NGC 5128, among
many other giant elliptical galaxies, is not in isolation, and
therefore not necessarily expected to reproduce such a low spin parameter.
Also, the internal rotation axis changes at 5 kpc are not easily explained with
only the monolithic collapse scenario.  In the monolithic collapse 
model, the inner regions would be expected
to have more pronounced rotations.
Yet all four of the galaxies with major kinematic studies presented
here do not show a higher rotational signal in the inner
regions. In fact, for NGC 1399, the outer region ($R > 6'$) indicates
rotation in the metal-poor population that is not evident in the inner
regions.  Also, a slightly lower rotational signal is present in the
inner regions of NGC 5128 for the metal-poor population than in the
outer regions.  

Hierarchical clustering of cold dark matter also 
relies on angular momentum in a galaxy being produced by
gravitational tidal torques during the growth of initial
perturbations.  \cite{sugerman00} have demonstrated that the 
tidal torque theory predicts an increase in
angular momentum during the collapse, and with time, the increase in
angular momentum slows.   Accretion of
satellites and/or merger events is therefore a possible culprit for moving the angular
momentum outward, as major mergers of disks and bulges 
suggest that angular momentum resides largely in the outer regions of the galaxy
\citep{barnes92,hernquist93}.  

Alternatively, \cite{vitvitska02} examine the change in spin parameter
in a scenario where the angular momentum in a galaxy is built
up by mass accretion.  Their results show that the spin parameter changes
sharply in major merger events in the galaxy and steadily
decreases with small satellite accretion events.  They also show
that the spin parameter for a galaxy with a major merger after a
redshift of $z = 3$ should be notably larger than a galaxy that
did not undergo such a major merger.  Their study obtains an average of
$\Lambda = 0.045$ from $\Lambda$CDM $N$-body simulations for galaxies
with halo masses of $(1.1-1.5) \times 10^{12} h^{-1}$ $M_{\odot}$ with
$h = 0.7$.

NGC 1399 and M49, with their weak rotation signals, are consistent
with the model predictions discussed above, whereby
their major formation events could have occurred at early times and with perhaps
only minor accretions happening since then.  However, NGC 5128 and M87 have
spin parameters 2-3 times larger than predicted by the model 
averages.  For NGC 5128, this relatively large rotation (which is nearly
independent of both metallicity and radius) may, perhaps, be connected
with its history within the Centaurus group environment.  The rotation speed
and rotation axis for its extended group of satellite galaxies are nearly identical to the
NGC 5128 halo \citep{w06}, much as if the accretion events 
experienced by the central giant have been taking place preferentially along the
main axis of the entire group and in the same orientation.  The GCS age
distribution discussed by \cite{beasley06} 
and the mean age for the halo field stars \citep{rejkuba05} strongly
suggest that a high fraction of the stellar population in NGC 5128 formed
long ago, with particularly large bursts between 8 and 12 Gyr.  Even if the galaxy underwent a significant merger 
perhaps a few Gyr ago (the traces of which now appear in the halo arcs
and shells), the stars in it may already have been old at the time
of the merger.  Although a very few younger GCs have formed
since then, these make up a small minority of what is present, at least
for the $R > 5$ kpc halo outside the bulge region that now contains gas and dust.

The situation for M87, with its even larger rotation signature, may
require a different sort of individual history.  Of the four galaxies 
compared here, it is at the dynamical center of the richest environment 
(Virgo), has the most extensive cD-type envelope, and sits within
the most massive, extended, and dynamically evolved potential well.
A single relatively recent major merger could in principle have caused
its present high rotation, but the lack of distinctive tidal features
does not necessarily favor such an interpretation and would at least
suggest that such a merger should have been with another large elliptical
and nondissipative galaxy.  \cite{cote01} discuss an interpretation - at
least partially resembling what we suggest for NGC 5128 - that stellar
material ''is gradually infalling onto M87 along the so-called principal
axis of the Virgo Cluster.''

In conclusion, we have presented a kinematic study of NGC 5128 
that makes it now comparable to recent
studies of the other giant galaxies, M87, M49, and NGC 1399.  
Using $340$ GCs ($158$ metal-rich and $178$
metal-poor GCs), we have calculated
the rotation amplitude, rotation axis, and velocity dispersion and
have searched for radial and metallicity dependences.  Our findings show that both
metallicity populations rotate with little dependence on projected
radius, with $\Omega R = 40\pm10$, 31$\pm$14, and 47$\pm$15 km s$^{-1}$ for the total, metal-poor, and
metal-rich populations, respectively.  Perhaps the inner 5 kpc shows a 
slower rotation of the
metal-poor population, but more clusters would be needed to confirm this
finding.  The rotation axis is 189$\pm$12$^o$, 177$\pm$22$^o$, 
and 202$\pm$15$^o$ east of north for the total, metal-poor, and
metal-rich populations out to a 50 kpc projected radius, assuming the
velocity field is best fit by a sine curve.  
The rotation axis does change at 5 kpc, following the zero-velocity 
curve proposed by \cite{peng04} or possibly full-on counterrotation.
A study with more GCs and lower uncertainties is needed to see what is
happening in the innermost 5 kpc of NGC 5128.  The velocity dispersion
shows a modest increase with galactocentric radius, although the outer
regions (especially the metal-rich population) have less reliable
statistics; this increase could be driven purely by statistical
effects.  We find the velocity dispersion we find  
123$\pm$5, 117$\pm$7, and 129$\pm$9 km s$^{-1}$ for the total, metal-poor, and
metal-rich populations, respectively. 

The PN data are also used to determine the kinematics of the halo of
NGC 5128.  These show results that are encouragingly similar to those of the GC
data, except that no rotation axis change is noted with radius, and a
{\it decrease} in velocity dispersion is found with radius, possibly
indicating a difference in orbital anisotropy compared with the
GCs.  A very similar effect has been noted for the Leo
elliptical NGC 3379, although with a much smaller data sample 
\citep[][]{romanowsky03,bergond06,pierce06}.  We also determine the total
dynamical mass using both the GCs and
the PNe by separately calculating the pressure supported
mass with the tracer mass estimator and the rotationally supported
mass using the spherical component of the Jeans equation.  The total
mass is $(1.3\pm0.5) \times 10^{12}$ $M_{\odot}$
from the GC population out to a projected radius
of 50 kpc, or $(1.0\pm0.2) \times 10^{12}$ $M_{\odot}$ out to 90
kpc from the PNe.  

Overall, we have enough evidence to cautiously
conclude that a major episode of star formation
occurred about $8-10$ Gyr ago (corresponding to
a redshift z = 1.2 - 1.8) and this may have been
when the bulk of the visible galaxy was built.
We still do not know just why the most
metal-poor clusters show up in such relatively large
numbers and appear to have ages of $10-12$ Gyr, but this is a common
issue in all big galaxies.

This kinematic study and the age study of \cite{beasley06} on the NGC 5128 cluster system indicate that
additional spectroscopic studies to build up both the radial velocity
database and age distribution can lead to rich dividends.  Large
GC samples are clearly needed to remove the current
sample biases and to 
fully understand the complex kinematics and history of this giant elliptical galaxy.
It seems clear as well that each galaxy needs to be individually studied to fully
understand the different galaxy formation histories.  We are
continuing these studies particularly for NGC 5128, with the eventual
aim of at least doubling the total GC sample size in
this unique system.

Acknowledgements: WEH and GLHH acknowledge financial support from
NSERC through operating research grants. DAF thanks the ARC for
financial support.

\clearpage
\pagestyle{empty}
\setlength{\voffset}{25mm}
\begin{deluxetable}{llllrcccccccccccccccccccccccl}
\rotate
\setlength{\tabcolsep}{0.02in}
\tabletypesize{\tiny}
\tablecaption{The Globular Cluster System Catalogue of NGC 5128 \label{tab:cat_GC}}
\tablewidth{0pt}
\tablehead{
\colhead{Name} &\colhead{Old Name} & \colhead{RA} & \colhead{Dec.} & \colhead{R} & \colhead{$U$} & \colhead{$B$} & \colhead{$V$} & \colhead{$R$} & \colhead{$I$} & \colhead{$\sigma_U$} & \colhead{$\sigma_B$} & \colhead{$\sigma_V$} & \colhead{$\sigma_R$} & \colhead{$\sigma_I$} & \colhead{$C$} & \colhead{$M$} & \colhead{$T_1$} & \colhead{$\sigma_C$} & \colhead{$\sigma_M$} & \colhead{$\sigma_{T_1}$} & \colhead{$U-B$} & \colhead{$B-V$} & \colhead{$V-R$} & \colhead{$V-I$} & \colhead{$M-T_1$} & \colhead{$C-M$} & \colhead{$C-T_1$} & \colhead{$v_r$} \\
\colhead{ } & \colhead{ } &\colhead{(J2000)} & \colhead{(J2000) } & \colhead{(')} & \colhead{(mag)} & \colhead{(mag)} & \colhead{(mag)} & \colhead{(mag)} & \colhead{(mag)} & \colhead{(mag)} & \colhead{(mag)}  & \colhead{(mag)} & \colhead{(mag)} & \colhead{(mag)} & \colhead{(mag)} & \colhead{(mag)} & \colhead{(mag)} & \colhead{(mag)} & \colhead{(mag)} & \colhead{(mag)} & \colhead{(mag)} & \colhead{(mag)} & \colhead{(mag)} & \colhead{(mag)} & \colhead{(mag)} & \colhead{(mag)} & \colhead{(mag)}  & \colhead{(km s$^{-1}$)}  \\
}

\startdata
GC0001 & pff\_gc-028  & 13 20 01.16 &-42 56 51.5&  6.46 & 20.11& 19.68& 18.75& 18.20& 17.60& 0.07& 0.02& 0.01& 0.01& 0.03&  19.92& 19.04& 18.22& 0.01& 0.01& 0.01& 0.43& 0.93& 0.55& 1.15& 0.82& 0.88& 1.70&   524$\pm$16\tablenotemark{5} \\  
GC0002 &  HH-048      & 13 22 45.36 &-43 07 08.8& 30.26 &   -  &   -  &   -  &   -  &   -  &  -  &  -  &  -  &  -  &  -  &    -  &   -  &   -  &  -  &  -  &  -  &  -  &  -  &  -  &  -  &  -  &  -  &  -  &   856$\pm$50\tablenotemark{7} \\ 
GC0003 &  HH-080      & 13 23 38.33 &-42 46 22.8& 24.84 &   -  &   -  &   -  &   -  &   -  &  -  &  -  &  -  &  -  &  -  &  19.45& 18.72& 17.98& 0.02& 0.01& 0.01&  -  &  -  &  -  &  -  & 0.74& 0.72& 1.46&   497$\pm$33\tablenotemark{7} \\ 
GC0004 & HGHH-40/HH-044/C40 & 13 23 42.37 &-43 09 37.8& 21.02 & 19.37& 19.01& 18.12& 17.59& 17.03& 0.04& 0.01& 0.01& 0.01& 0.03&  20.12& 19.28& 18.49& 0.02& 0.01& 0.01& 0.35& 0.89& 0.53& 1.10& 0.79& 0.84& 1.63&   365$\pm$24\tablenotemark{4,5,7} \\ 
GC0005 & HGHH-01/C1   & 13 23 44.19 &-43 11 11.8& 21.41 &   -  &   -  &   -  &   -  &  -   &  -  &  -  &  -  &  -  &  -  &  18.73& 18.00& 17.25& 0.06& 0.03& 0.03&  -  &  -  &  -  &  -  & 0.76& 0.72& 1.48&   642$\pm$1\tablenotemark{1,5,7,8} \\   
GC0006 & pff\_gc-001  & 13 23 49.62 &-43 14 32.0& 22.36 & 20.39& 19.91& 18.91& 18.36& 17.60& 0.09& 0.02& 0.01& 0.01& 0.03&  20.22& 19.25& 18.38& 0.02& 0.01& 0.01& 0.47& 1.01& 0.55& 1.31& 0.87& 0.97& 1.84&   711$\pm$35\tablenotemark{5,7} \\ 
GC0007 & AAT301956    & 13 23 54.52 &-43 20 01.1& 25.41 & 21.22& 21.13& 20.42& 19.98& 19.43& 0.18& 0.06& 0.02& 0.02& 0.04&    -  &   -  &   -  &  -  &  -  &  -  & 0.09& 0.71& 0.44& 0.98&  -  &  -  &  -  &   287$\pm$162\tablenotemark{7} \\ 
GC0008 &  HH-099      & 13 23 56.70 &-42 59 59.8& 16.66 &   -  &   -  &   -  &   -  &   -  &  -  &  -  &  -  &  -  &  -  &  21.72& 20.90& 20.01& 0.09& 0.09& 0.06&  -  &  -  &  -  &  -  & 0.89& 0.82& 1.71&   798$\pm$49\tablenotemark{7} \\
\enddata																														      
\tablenotetext{1}{Radial velocity measured by \cite{vhh81}.}
\tablenotetext{2}{Radial velocity measured by \cite{hhvh84}.}
\tablenotetext{3}{Radial velocity measured by \cite{hhh86}.}
\tablenotetext{4}{Radial velocity measured by \cite{hghh92}.}
\tablenotetext{5}{Radial velocity measured by \cite{pff04I}.}
\tablenotetext{6}{Radial velocity measured by \cite{whh05}.}
\tablenotetext{7}{Radial velocity measured by \cite{beasley06}.}
\tablenotetext{8}{Radial velocity measured by \cite{rejkuba07}.}
\tablenotetext{9}{These objects appear starlike under HST/ACS imaging \citep{harris06}.}
\tablenotetext{10}{This object appears to be a M-type star based on the strong
 molecular bands in its spectrum \citep{beasley06}.}
\tablenotetext{11}{Faint globular \citep{harris06}.}
\tablenotetext{12}{Located in the inner bulge region which is very
  crowded \citep{harris06}. }
\tablenotetext{13}{A very compact globular or star \citep{harris06}.}
\tablenotetext{14}{A possible globular, but slightly elliptical \citep{harris06}.}
\tablenotetext{15}{A globular next to a very bright star \citep{harris06}.}

\end{deluxetable}	   

\clearpage
\pagestyle{plaintop}
\setlength{\voffset}{0mm}

\begin{deluxetable}{rrrrrrrrrrr}
\tabletypesize{\scriptsize}
\tablecaption{Kinematic and Dynamic Solutions for the Globular Cluster System of NGC 5128\tablenotemark{1} \label{tab:all_GC}}
\tablewidth{0pt}
\tablehead{
\colhead{Radial Bin} & {R$_{mean}$}  & {R$_{out}$} & \colhead{N} & \colhead{$\Omega
  R$} & \colhead{$\Theta_o$} & \colhead{$\sigma_v$} &
\colhead{$M_{corr}$} & \colhead{$M_p$} & \colhead{$M_r$} &
\colhead{$M_t$} \\
\colhead{(kpc)} & \colhead{(kpc)} & \colhead{(kpc)} & \colhead{ } & \colhead{(km
  s$^{-1}$)} & \colhead{($^o$ E of N)} & \colhead{(km s$^{-1}$)} &
\colhead{ } & \colhead{($\times 10^{10} M_{\odot}$)} & \colhead{($\times 10^{10}
  M_{\odot}$)} & \colhead{($\times 10^{10} M_{\odot}$)}  \\
}

\startdata
0-50  &	12.9 & 48.8 & 340 &  40$\pm$10 &  189$\pm$12  & 123$\pm$5  & 1.0 & 125.8$\pm$46.5\tablenotemark{2} & 3.1$\pm$1.3  & 128.9$\pm$46.5  \\
0-5   &	3.64 & 4.96 &  54 &  24$\pm$21 &  334$\pm$59  & 120$\pm$12 &  -  &   -            &    -         &  -              \\
5-10  &	7.65 & 9.96 & 124 &  43$\pm$15 &  195$\pm$20  & 112$\pm$8  & 2.3 &  37.4$\pm$6.4  &  0.4$\pm$0.3 &  37.8$\pm$6.4  \\
10-15 &	12.4 & 14.9 &  68 &  83$\pm$25 &  195$\pm$12  & 105$\pm$10 & 1.6 &  36.5$\pm$9.3  &  2.4$\pm$1.5 &  38.9$\pm$9.4  \\
15-25 & 19.0 & 24.3 &  56 &  35$\pm$26 &  184$\pm$34  & 147$\pm$16 & 1.3 &  89.5$\pm$27.4 &  0.7$\pm$1.0 &  90.2$\pm$27.4 \\ 
25-50 &	34.7 & 48.8 &  39 &  96$\pm$45 &  169$\pm$17  & 148$\pm$21 & 1.0 & 184.7$\pm$84.5 & 10.4$\pm$9.8 & 195.1$\pm$85.1 \\ 
\enddata

\tablenotetext{1}{This table is reproduced from \cite{w06} for
  completeness.}
\tablenotetext{2}{The Tracer Mass estimator determines the total mass
  enclosed within the outermost radius using a sample defined in the
  range of the inner and outer radial bins.}

\end{deluxetable}

\begin{deluxetable}{rrrrrrrrrrr}
\tabletypesize{\scriptsize}
\tablecaption{Kinematic and Dynamic Solutions for the Metal-Poor Globular Cluster System of NGC 5128 \label{tab:MP_GC}}
\tablewidth{0pt}
\tablehead{
\colhead{Radial Bin} & {R$_{mean}$} & {R$_{out}$} & \colhead{N} & \colhead{$\Omega
  R$} & \colhead{$\Theta_o$} & \colhead{$\sigma_v$} &
\colhead{$M_{corr}$} & \colhead{$M_p$} & \colhead{$M_r$} &
\colhead{$M_t$} \\
\colhead{(kpc)} & \colhead{(kpc)}  & \colhead{(kpc)} & \colhead{ } & \colhead{(km
  s$^{-1}$)} & \colhead{($^o$ E of N)} & \colhead{(km s$^{-1}$)} &
\colhead{ } & \colhead{($\times 10^{10} M_{\odot}$)} & \colhead{($\times 10^{10}
  M_{\odot}$)} & \colhead{($\times 10^{10} M_{\odot}$)}  \\
}

\startdata
0-50  & 13.8 & 48.1 &  178 &   31$\pm$14 &  177$\pm$22 & 117$\pm$7   & 1.0 & 94.5$\pm$58.5  &  1.1$\pm$1.0 &  95.6 $\pm$58.6  \\
0-5   & 3.84 & 4.96 &   22 &   16$\pm$38 &  373$\pm$120&  99$\pm$19  & -   &   -            &        -     &       -          \\    
5-10  & 7.76 & 9.96 &   66 &   30$\pm$21 &  223$\pm$39 & 116$\pm$12  & 2.1 &  38.9$\pm$12.4 & 0.2$\pm$0.3  &  39.1$\pm$12.4  \\
10-15 & 12.7 & 14.8 &   41 &   97$\pm$38 &  199$\pm$13 & 102$\pm$15  & 1.5 &  31.4$\pm$15.1 & 3.3$\pm$2.5  &  34.6$\pm$15.3  \\
15-25 & 18.8 & 24.3 &   26 &   45$\pm$31 &  139$\pm$49 & 112$\pm$20  & 1.3 &  59.7$\pm$32.7 & 1.2$\pm$1.6  &  60.8$\pm$32.7  \\
25-50 & 36.7 & 48.1 &   23 &   61$\pm$43 &  137$\pm$50 & 141$\pm$24  & 1.0 & 122.4$\pm$82.9 & 4.2$\pm$5.9  & 126.7$\pm$83.1  \\
\enddata

\end{deluxetable}

\begin{deluxetable}{rrrrrrrrrrr}
\tabletypesize{\scriptsize}
\tablecaption{Kinematic and Dynamic Solutions for the Metal-Rich Globular Cluster System of NGC 5128 \label{tab:MR_GC}}
\tablewidth{0pt}
\tablehead{
\colhead{Radial Bin} & {R$_{mean}$} & {R$_{out}$}& \colhead{N} & \colhead{$\Omega
  R$} & \colhead{$\Theta_o$} & \colhead{$\sigma_v$} &
\colhead{$M_{corr}$} & \colhead{$M_p$} & \colhead{$M_r$} &
\colhead{$M_t$} \\
\colhead{(kpc)} & \colhead{(kpc)} & \colhead{(kpc)} & \colhead{ } & \colhead{(km
  s$^{-1}$)} & \colhead{($^o$ E of N)} & \colhead{(km s$^{-1}$)} &
\colhead{ } & \colhead{($\times 10^{10} M_{\odot}$)} & \colhead{($\times 10^{10}
  M_{\odot}$)} & \colhead{($\times 10^{10} M_{\odot}$)}  \\
}

\startdata
0-50  &	12.0 & 48.8 &  158  &   47$\pm$15 &  202$\pm$15 & 129$\pm$9  & 1.0 &   116.5$\pm$73.8  &   4.3$\pm$2.3  &   120.8$\pm$73.9  \\
0-5   &	3.61 & 4.98 &   29  &   44$\pm$27 &  308$\pm$47 & 134$\pm$23 &  -  &     -             &        -       &          -        \\ 
5-10  & 7.52 & 9.93 &   58  &   65$\pm$22 &  180$\pm$19 & 105$\pm$11 & 1.5 &   17.4$\pm$7.7    &   1.0$\pm$0.7  &    18.3$\pm$7.7   \\
10-15 &	12.0 & 14.9 &   27  &   74$\pm$34 &  186$\pm$22 & 108$\pm$17 & 1.2 &   23.8$\pm$11.3   &   1.9$\pm$1.8  &    25.7$\pm$13.3  \\
15-25 & 19.3 & 24.3 &   30  &   50$\pm$36 &  223$\pm$32 & 168$\pm$24 & 1.1 &   72.1$\pm$46.2   &   1.2$\pm$1.7  &    73.3$\pm$46.3  \\
25-50 &	31.7 & 48.8 &   15  &  102$\pm$96 &  191$\pm$21 & 146$\pm$65 & 1.0 &  219.1$\pm$174.6  &  11.8$\pm$22.2 &   230.9$\pm$176.0 \\
\enddata

\end{deluxetable}

\begin{deluxetable}{rrrrrrrrrrr}
\tabletypesize{\scriptsize}
\tablecaption{Kinematic and Dynamic Solutions for the Planetary Nebula System of NGC 5128 \label{tab:PN}}
\tablewidth{0pt}
\tablehead{
\colhead{Radial Bin} & {R$_{mean}$} & {R$_{out}$} & \colhead{N} & \colhead{$\Omega
  R$} & \colhead{$\Theta_o$} & \colhead{$\sigma_v$} & \colhead{$M_{corr}$} & \colhead{$M_p$} & \colhead{$M_r$} &
\colhead{$M_t$} \\
\colhead{(kpc)} & \colhead{(kpc)} & \colhead{(kpc)}& \colhead{ } & \colhead{(km
  s$^{-1}$)} & \colhead{($^o$ E of N)} & \colhead{(km s$^{-1}$)}  &
\colhead{ } & \colhead{($\times 10^{10} M_{\odot}$)} & \colhead{($\times 10^{10}
  M_{\odot}$)} & \colhead{($\times 10^{10} M_{\odot}$)} \\
}

\startdata
0-90  &14.1 & 88.2 & 780 &   76$\pm$6  &  170$\pm$5  & 118$\pm$13 & 1.0 & 84.6$\pm$17.2  &11.9$\pm$1.9   & 96.5$\pm$17.3  \\
0-5   & 3.3 & 4.99 & 184 &   75$\pm$16 &  179$\pm$11 & 131$\pm$7  &  -  &      -         &     -         &     -          \\
5-10  & 7.6 & 9.98 & 211 &   82$\pm$12 &  177$\pm$8  & 120$\pm$6  & 3.0 & 43.9$\pm$3.6   & 1.6$\pm$0.5   & 45.4$\pm$3.6  \\
10-15 &11.8 & 14.8 & 138 &   76$\pm$15 &  171$\pm$10 & 118$\pm$7  & 2.0 & 42.3$\pm$5.0   & 2.0$\pm$0.8   & 44.2$\pm$5.1  \\
15-20 &17.4 & 20.0 &  71 &   96$\pm$24 &  169$\pm$12 & 116$\pm$10 & 1.7 & 47.7$\pm$5.6   & 4.3$\pm$2.1   & 52.0$\pm$6.9  \\
20-30 &24.6 & 30.0 &  87 &   76$\pm$17 &  157$\pm$17 & 108$\pm$8  & 1.5 & 53.2$\pm$8.4   & 4.0$\pm$1.8   & 57.2$\pm$8.6  \\
30-40 &34.6 & 39.8 &  50 &   44$\pm$15 &  132$\pm$38 &  87$\pm$9  & 1.3 & 40.7$\pm$7.4   & 1.8$\pm$1.2   & 42.3$\pm$7.5  \\
40-80 &48.7 & 71.2 &  36 &   61$\pm$45 &  183$\pm$22 &  85$\pm$11 & 1.0 & 48.6$\pm$11.5  & 6.2$\pm$9.1   & 54.8$\pm$14.7 \\
\enddata

\end{deluxetable}

\begin{deluxetable}{lrrrrrrrr}
\tabletypesize{\footnotesize}
\tablecaption{Spin Parameters for Giant Elliptical Galaxies \label{tab:spin}}
\tablewidth{0pt}
\tablehead{
\colhead{Galaxy} & \colhead{$\Omega R_{MP}$} & \colhead{$\sigma_{v,MP}$} & \colhead{$(\Omega R / \sigma_v)_{MP}$} & \colhead{$\Lambda_{MP}$} & \colhead{$\Omega R_{MR}$} & \colhead{$\sigma_{v,MR}$} & \colhead{$(\Omega R / \sigma_v)_{MR}$} & \colhead{$\Lambda_{MR}$} \\
\colhead{ } & \colhead{(km s$^{-1}$)} & \colhead{(km s$^{-1}$)}& \colhead{ } &\colhead{ } & \colhead{(km s$^{-1}$)} & \colhead{(km s$^{-1}$)}& \colhead{ } & \colhead{ } \\
}

\startdata 
NGC 5128 & 31$\pm$14 & 117$\pm$7 & 0.26$\pm$0.12 & 0.08$\pm$0.04 & 47$\pm$15 & 129$\pm$9 & 0.36$\pm$0.11 & 0.11$\pm$0.03 \\
M87\tablenotemark{8,9}    &186$^{+58}_{-41}$ &397$^{+36}_{-14}$ & 0.47$^{+0.13}_{-0.11}$& 0.14$^{+0.04}_{-0.03}$ &155$^{+53}_{-37}$ &365$^{+38}_{-18}$ & 0.43$^{+0.14}_{-0.12}$ & 0.13$^{+0.04}_{-0.04}$ \\
M49\tablenotemark{9}     & 93$^{+69}_{-37}$ &342$^{+33}_{-18}$ & 0.27$^{+0.19}_{-0.11}$& 0.08$^{+0.06}_{-0.03}$ &-26$^{+64}_{-79}$ &265$^{+34}_{-13}$ & 0.10$^{+0.27}_{-0.25}$ & 0.03$^{+0.08}_{-0.08}$ \\    
NGC 1399\tablenotemark{10} & 15$\pm$26 &291$\pm$14 & 0.05$\pm$0.09 & 0.02$\pm$0.03 & 7$\pm$24 & 255$\pm$13 & 0.03$\pm$0.09 & 0.01$\pm$0.03 \\
\enddata

\tablenotetext{8}{Kinematic data taken from \cite{cote01}.}
\tablenotetext{9}{Kinematic data taken from \cite{cote03}.}
\tablenotetext{10}{Kinematic data taken from \cite{richtler04}.}
\end{deluxetable}

\clearpage

\clearpage

\clearpage

\clearpage

\clearpage

\clearpage

\clearpage

\clearpage

\clearpage

\clearpage

\clearpage

\clearpage

\clearpage

\clearpage

\clearpage

\begin{thebibliography}{}
\bibitem[Arnoaboldi et al.(1994)]{arnaboldi94} Arnaboldi, M., Freeman,
  K.C., Capaccioli, M., \& Ford, H. 1994, ESO Messanger, 76, 40
\bibitem[Arnaboldi et al.(1998)]{arnaboldi98} Arnaboldi, M., Freeman,
  K.C., Gerhard, O., Matthias, M., Kudritzki, R.P. M\'endez, R.H.,
  Capaccioli, M., \& Ford, H. 1998, \apj, 507, 759
\bibitem[Baade \& Minkowski(1954)]{baade54} Baade, W., \& Minkowski,
  R. 1954, \apj, 119, 215
\bibitem[Bahcall \& Tremaine(1981)]{bahcall81} Bahcall, J.N., \&
  Tremaine, S. 1981, \apj, 244, 805
\bibitem[Barnes(1992)]{barnes92} Barnes, J.E. 1992, \apj, 393, 484
\bibitem[Beasley et al.(2004)]{beasley04} Beasley, M.A., Forbes, D.A.,
  Brodie, J.P., \& Kissler-Patig, M. 2004, \mnras, 347, 1150
\bibitem[Beasley et al.(2006)]{beasley06} Beasley, M.A., Bridges, T.,
  Peng, E., Harris, W.E., Harris, G.L.H., Forbes, D.A., \& Mackie,
  G. 2006, in preparation 
\bibitem[Bekki, Harris, \& Harris(2003)]{bhh03} Bekki, K., Harris,
  W.E., \& Harris, G.L.H. 2003, \mnras, 338, 587
\bibitem[Bekki et al.(2003)]{bekki03} Bekki, K., Forbes, D.A.,
  Beasley, M.A., \& Couch, W.J. 2003, \mnras, 334, 1334
\bibitem[Bekki \& Peng(2006)]{bekki06} Bekki, K., \& Peng, E.W. 2006,
  \mnras, 370, 1737
\bibitem[Bergond et al.(2006)]{bergond06} Bergond, G., Zepf, S.E.,
  Romanowsky, A.J., Sharples, R.M., \& Rhode, K.L. 2006, \aap, 448, 155
 \bibitem[Bridges et al.(2006)]{bridges06} Bridges, T., Gebhardt, K.,
  Sharples, R., Faifer, F.R., Forte, J.C., Beasley, M.A., Zepf, Z.E.,
  Forbes, D.A., Pierce, M. 2006, \mnras, 373, 157
\bibitem[Ciardullo et al.(1993)]{ciardullo93} Ciardullo, R., Jacoby,
  G.H., \& Dejonghe, H.B. 1993, \apj, 414, 454
\bibitem[C\^ot\'e et al.(2001)]{cote01} C\^ot\'e, P., McLaughlin, D.E., Hanes,
  D.A., Bridges, T.J., Geisler, D., Merritt, D., Hesser, J.E., Harris,
  G.L.H., \& Lee, M.G. 2001, \apj, 559, 828
\bibitem[C\^ot\'e et al.(2003)]{cote03} C\^ot\'e, P., McLaughlin, D.,
  Cohen, J.G., \& Blakesless, J.P. 2003, \apj, 591, 850
\bibitem[Dekel et al.(2005)]{dekel05} Dekel, A., Stoehr, F., Mamon,
  G.A., Cox, T.J., Novak, G.S., \& Primak, J.R., 2005, Nature, 437, 707
\bibitem[Dufour et al.(1979)]{dufour79} Dufour, R.J., Harvel, C.A.,
  Martins, D.M., Schiffer, F.H., Talent, D.L., Wells, D.C., van den
  Bergh, S., \& Talbot, R.J. 1979, \aj, 84, 284
\bibitem[Efstathiou \& Jones(1979)]{efstathiou79} Efstathiou, G., \&
  Jones, B.J.T. 1979, \mnras, 186, 133
\bibitem[Evans et al.(2003)]{evans03} Evans, N.W., Wilkinson, M.I.,
  Perrett, K.M., \& Bridges, T.J. 2003, \apj, 583, 752
\bibitem[Fall(1979)]{fall79} Fall, S.M. 1979, Rev. Mod. Phys., 51, 21
\bibitem[Ferrarese et al.(2006)]{ferr06} Ferrarese, L., Mould, J.R.,
 Stetson, P.B., Tonry, J.L., Blakeslee, J.P., \& Ajhar, E.A. 2006,
 ApJ, in press (astro-ph/0605707)
\bibitem[Harris et al.(1992)]{hghh92} Harris, G.L.H., Geisler, D., Harris, H.C., \& Hesser, J.E. 1992, \aj, 104, 613
\bibitem[Harris et al.(1999)]{hhp99} Harris, G.L.H., Harris, W.E., \& Poole, G. 1999, AJ, 117, 855
\bibitem[Harris \& Harris(2002)]{harris02} Harris, W. E., \& Harris,
  G. L. H. 2002, \aj, 123, 3108
\bibitem[Harris et al.(2002)]{hhhm02} Harris, W.E., Harris, G.L.H.,
  Holland, S.T., \& McLaughlin, D.E. 2002, \aj, 124, 1435
\bibitem[Harris et al.(2004)]{hhg04II} Harris, G.L.H., Harris, W.E.,
  \& Geisler, D. 2004, \aj, 128, 723
\bibitem[Harris et al.(2006)]{harris06} Harris, W.E., Harris, G.L.H.,
  Barmby, P., McLaughlin, D.E., \& Forbes, D. 2006, \aj, 132, 2187
\bibitem[Heisler, Tremaine, \& Bahcall(1985)]{heisler85} Heisler, J.,
  Tremaine, S., \& Bahcall, J.N. 1985, \apj, 298, 8
\bibitem[Hernquist(1993)]{hernquist93} Hernquist, I. 1993, \apj, 409, 548
\bibitem[Hesser et al.(1984)]{hhvh84} Hesser, J.E., Harris, H.C., van den Bergh, S., \& Harris, G.L.H. 1984, \apj, 276,491
\bibitem[Hesser, Harris, \& Harris(1986)]{hhh86} Hesser, J.E., Harris,
  H.C., \& Harris, G.L.H. 1986, \apj, 303, L51
\bibitem[Hui et al.(1995)]{hui95} Hui, X., Ford, H.C., Freeman, K.C., \&
  Dopita, M.A. 1995, \apj, 449, 592
\bibitem[Kraft et al.(2003)]{kraft03} Kraft, R.P., V\'aquez, S.E.,
  Forman, W.R., Jones, C., Murray, S.S., Hardcastle, M.J., Worrall,
  D.M., Churazov, E. 2003, \apj, 592, 129
\bibitem[Larsen et al.(2002)]{larsen02} Larsen, S.S., Brodie, J.P.,
  Beasley, M.A., \& Forbes, D.A. 2002, \aj, 124, 828
\bibitem[Lee \& Worthey(2005)]{lee05} Lee, H., \& Worthey, G. 2005,
  \apjs, 160, 176
\bibitem[Ma\'iz Apell\'aniz \& \'Ubeda(2005)]{maiz05} Ma\'iz Apell\'aniz, J. \& \'Ubeda,
  L. 2005, \apj, 629, 7873
\bibitem[Mamon \& {\L}okas(2005)]{mamon05} Mamon, G.A., \& {\L}okas,
  E.L. 2005, \mnras, 362, 95
\bibitem[Napolitano et al.(2001)]{napolitano01} Napolitano, N.R.,
  Arnaboldi, M., Freeman, K.C., \& Capaccioli, M. 2001, \aap, 377, 784
\bibitem[Napolitano et al.(2002)]{napolitano02} Napolitano, N.R.,
  Arnaboldi, M., \& Capaccioli, M. 2002, \aap, 383, 791
\bibitem[O'Sullivan, Forbes, \& Ponman(2001)]{osullivan01} O'Sullivan, E., Forbes,
  D.A., \& Ponman, T.J. 2001,\mnras, 328, 461 
\bibitem[Peebles(1969)]{peebles69} Peebles, P.J.E. 1969, \apj, 155, 393
\bibitem[Peng et al.(2002)]{peng02} Peng, E.W., Ford, H.C., Freeman, K.C., \& White, R.L. 
2002, \aj, 124, 3144
\bibitem[Peng, Ford, \& Freeman(2004a)]{pff04I} Peng, E.W., Ford, H.C.,
  \& Freeman, K.C. 2004, \apjs, 150, 367
\bibitem[Peng, Ford \& Freeman(2004b)]{peng04} Peng, E.W., Ford, H.C., \&
  Freeman, K. 2004, \apj, 602, 685
\bibitem[Peng, Ford \& Freeman(2004c)]{pff04II} Peng, E.W., Ford, H.C., \&
  Freeman, K. 2004, \apj, 602, 705
\bibitem[Pierce et al.(2006)]{pierce06} Pierce, M., Beasley, M.A.,
  Forbes, D.A., Bridges, T., Gebhardt, K., Faifer, F.R., Forte, J.C.,
  Zepf, S.E., Sharples, R., Hanes, D.A., \& Proctor, R. 2006, \mnras,
  366, 1253
\bibitem[Press et al.(1992)]{press92} Press, W.H., Teukolsky, S.A.,
  Vetterling, W.T., \& Flannery, B.P. 1992, Numerical Recipes in
  Fortran (2nd Ed., Cambridge:  Camdridge University Press)
\bibitem[Puzia et al.(2005)]{puzia05} Puzia, T.H., Kissler-Patig, M.,
  Thomas, D., Maraston, C., Saglia, R.P., Bender, R., Goudfrooij, P.,
  \& Hempel, M. 2005, \aap, 439, 997 
\bibitem[Reed, Harris, \& Harris(1994)]{reed94} Reed, L.G., Harris,
  G.L.H., \& Harris, W.E. 1994, \aj, 107, 555
\bibitem[Rejkuba (2001)]{rejkuba01} Rejkuba, M. 2001, \aap, 269, 812
\bibitem[Rejkuba (2004)]{rejkuba04} Rejkuba, M. 2004, \aap, 413, 903
\bibitem[Rejkuba et al.(2005)]{rejkuba05} Rejkuba, M., Greggio, L.,
  Harris, W.E., Harris, G.L.H., \& Peng, E.W. 2005, \apj, 631, 262
\bibitem[Rejkuba et al.(2007)]{rejkuba07} Rejkuba, M., Dubath, P., 
  Minniti, D., \& Meylan, G. 2007, \aap, in press (astro-ph/0703385) 
\bibitem[Richtler et al.(2004)]{richtler04} Richtler, T., Dirsch, B.,
	Gebhardt, K., Geisler, D., Hilker, M., Alonso, M.~V., Forte,
	J.~C,. and Grebel, E.~K., Infante, L., Larsen, S., Minniti,
	D., \& Rejkuba, M. 2004, \aj, 127, 2094
\bibitem[Romanowsky et al.(2003)]{romanowsky03} Romanowsky, A.J.,
  Douglas, N.G., Arnaboldi, M., Kuijken, K., Merrifield, M.R.,
  Napolitano, N.R., Capaccioli, M., \& Freeman, K.C., 2003, Science, 301, 1696
\bibitem[Saglia et al.(2000)]{saglia00} Saglia, R.P., Kronawitter, A.,
  Gerhard, O., \& Bender, R. 2000, \aj, 119, 153
\bibitem[Samurovi\'c(2006)]{samurovic06} Samurovi\'c, S. 2006,
  Serb. Astron. J., 173, 35
\bibitem[Schiminovich et al.(1994)]{schiminovich94} Schiminovich, D.,
  van Gorkom, J.H., van der Hulst, J.M., \& Kasow, S. 1994, \apj, 423, L101
\bibitem[Sharples (1988)]{sharples88} Sharples, R. 1988, IAU
  Symp. 126:  The Harlow-Sharpley Symposium on Globular Cluster
  Systems in Galaxies, 126, 245
\bibitem[Strader et al.(2005)]{strader05} Strader, J., Brodie, J.P.,
  Cenarro, A.J., Beasley, M.A., \& Forbes, D.A. 2005, \aj, 130, 1315
\bibitem[Sugerman, Summers, \& Kamionkowski(2000)]{sugerman00}
  Sugerman, B., Summers, F.J., \& Kamionkowski, M. 2000, \mnras, 311, 762
\bibitem[Thomas, Maraston, \& Korn(2004)]{thomas04} Thomas, D.,
  Maraston, C., \& Korn, A. 2004, \mnras, 351, 19
\bibitem[Toomre \& Toomre(1972)]{toomre72} Toomre, A., \& Toomre,
  J. 1972, \apj, 178, 623
\bibitem[van den Bergh, Hesser, \& Harris(1981)]{vhh81} van den Bergh,
  S., Hesser, J.E., \& Harris, G.L.H. 1981, \aj, 86, 24
\bibitem[Vitvitska et al.(2002)]{vitvitska02} Vitvitska, M., Klypin,
  A.A., Kravtsov, A.V., Wechsler, R.H., Primack, J.R., \& Bullock,
  J.S. 2002, \apj, 581, 799
\bibitem[Woodley, Harris, \& Harris(2005)]{whh05} Woodley, K.A., Harris,
  W.E., \& Harris, G.L.H. 2005, \aj, 129, 2654
\bibitem[Woodley (2006)]{w06} Woodley, K.A. 2006, \aj, 132, 2424
\end{thebibliography}
\end{document}